\documentclass[hidelinks,onefignum,onetabnum]{siamart250211}

\usepackage{amsmath, amssymb, bm,thmtools}

\usepackage{multirow}
\usepackage{algorithm}
\usepackage[noend]{algorithmic} 

\usepackage{color}

\usepackage{soul}
\usepackage{subcaption}
\usepackage{tikz-network} 
\usepackage{braket}
\usepackage{pifont}
\usepackage{graphicx}

\usepackage{pgfplots}
\pgfplotsset{compat=1.17}

\usepackage{pdflscape}

\usepackage[T2A]{fontenc}

\newsiamremark{remark}{Remark}

\allowdisplaybreaks

\headers{A VQE-based cutting plane framework for SDP problems}{G. Ozbaygin, B. Kocuk, and D. A. Moran R.}

\title{A variational quantum eigensolver-based cutting plane framework for semidefinite programming problems\thanks{Submitted to the editors on 31 August 2026.
\funding{This work was funded by the Scientific and Technological Research Council of Turkey under grant number 125M579.}
}
}

\author{Gizem Ozbaygin\thanks{Industrial Engineering Department,  Bilkent University, Ankara, Turkey 06800 
  (\email{ozbaygin@bilkent.edu.tr}).} \and Burak Kocuk\thanks{Industrial Engineering Program,  Sabanc{\i} University, Istanbul, Turkey 34956 
  (\email{burakkocuk@sabanciuniv.edu}).}
\and Diego A. Moran R.\thanks{Industrial and Systems Engineering Department,  Rensselaer Polytechnic Institute, Troy, NY USA 12180-3590 
  (\email{morand@rpi.edu}).} }

\usepackage{amsopn}

\begin{document}

\maketitle

\begin{abstract}
Semidefinite programming plays a key role in optimization, with broad impact across control theory, machine learning, and combinatorial optimization. Although semidefinite programs are polynomially solvable, several commonly used algorithms rest on a linear-algebraic step whose running time grows cubically with the matrix dimension and which requires the matrix itself to be held in memory, at quadratic cost. In this study, we propose replacing it with a variational quantum eigensolver, whose qubit requirement is logarithmic in the matrix dimension, and present the first end-to-end implementation of such an approach within a cutting-plane framework, together with an operator-derived ansatz whose entanglement structure is read directly from the Pauli support of the candidate matrix. Evaluated on the \texttt{control} family of SDPLIB against an identical scheme driven by an exact eigendecomposition, the variational oracle produces valid cuts throughout, closing $32$ to $82\%$ of the initial optimality gap against a near-constant $75$ to $82\%$ for the exact oracle. Implementing and measuring the method end to end surfaces several effects not visible from theoretical analyses alone: where memory is actually consumed, how the padding required to fit a matrix onto a quantum register can mislead the variational optimizer, and why the candidate matrices prove dense in the Pauli basis, reducing the operator-derived ansatz to full entanglement. We report these findings and discuss their implications for near-term hybrid quantum-classical approaches.
\end{abstract}



\section{Introduction}
A semidefinite program (SDP) is an optimization problem of the following form:
\begin{equation} \label{eq:sdp-generic}
    \sup \; \{\, A_0 \bullet X \;:\; A_k \bullet X = b_k,\; k=1,\dots,m, \;\; X \in \mathbb{S}^n_+ \,\}
\end{equation}
Here, we have $A_0,\dots, A_m \in \mathbb{S}^n$,  $b_1,\dots,b_m \in \mathbb{R}$, where  $\mathbb{R}^n$, $\mathbb{S}^n$ and $\mathbb{S}_+^n$ respectively denote 
    the set of real $n$-vectors,
    the set of real symmetric $n\times n$ matrices and
    the set of real symmetric positive semidefinite (PSD) $n\times n$ matrices, and $X \bullet Y$ is the  Frobenius inner product between $X \in \mathbb{S}^n$ and $Y \in \mathbb{S}^n$.
SDPs serve as a cornerstone in convex programming, and conic programming in particular (see, classical references~\cite{todd2001semidefinite,vandenberghe1996semidefinite,wolkowicz2012handbook}). 
The use of
SDPs has lead to breakthroughs in  
combinatorial optimization~\cite{lovasz2003semidefinite,tunccel2016polyhedral},
graph theory~\cite{Goemans1995,pena2007computing}
polynomial optimization~\cite{lasserre2001global,parrilo2003semidefinite}.
More recently, SDPs have been heavily utilized in various problems domains including power systems~\cite{bai2008semidefinite}, signal processing~\cite{luo2010semidefinite}, wireless networks~\cite{liu2022integrated}, routing~\cite{de2009semidefinite}, quantum information theory~\cite{skrzypczyk2023semidefinite} among others.


While polynomially solvable via interior-point methods~\cite{nesterov1994interior}, large-scale SDPs are memory-intensive and challenging to solve computationally using these methods. Cutting plane  based methods are a promising alternative, especially for large-scale instances (see, \cite{krishnan2006unifying} for a review). In this paper, we propose a polyhedral outer-approximation cutting-plane algorithm: the starting point is a relaxation of the PSD constraints, and at each iteration we obtain a progressively tighter approximation of~\eqref{eq:sdp-generic} by introducing linear and second-order-cone valid inequalities, building on the polyhedral approximations of~\cite{wang2021polyhedral}.
An important step in cutting plane methods for SDPs is the PSD check of the current iterate, for which eigendecomposition is typically used. While this is a well-known task in classical linear algebra, it is also the step for which a quantum subroutine is most readily substituted: the variational quantum eigensolver (VQE) estimates the minimum eigenvalue of an $N \times N$ operator using a register of only $\lceil \log_2 N \rceil$ qubits, against the $O(N^2)$ numbers required to store the operator classically. Whether this logarithmic qubit requirement translates into an advantage inside a cutting-plane scheme is an empirical question, and it is mainly the question this paper aims to address.

We propose a cutting-plane framework in which the separation oracle is realized by VQE, and we study it computationally on the \texttt{control} family of SDPLIB against an identical scheme driven by an exact eigendecomposition, so that the effect of the oracle is isolated from that of the relaxation. Our contributions are as follows.

\begin{itemize}
\item We give a cutting-plane scheme with a variational separation oracle, together with two constructions: an \emph{operator-derived ansatz} whose entanglement map is induced by the Pauli support of the candidate matrix, and a \emph{multi-cut collection} strategy that extracts several violating directions from a single optimizer trajectory at no additional circuit cost.
\item We characterize the oracle against exact eigendecomposition at every cutting-plane iteration across ten instances, and identify a failure mode induced by the power-of-two embedding that causes the optimized state to collapse into the padded subspace, and falsely certifies a block PSD. We explain the mechanism, propose a remedy that removes it, and report its impact on the bounds returned by the algorithm.
\item We quantify the resource costs the logarithmic qubit count does not capture: (1) the memory bottleneck is measured to be the classical master problem rather than separation, (2) the candidate matrices are dense in the Pauli basis, and (3) cut generation requires the amplitudes of the optimized state, which on hardware would need state tomography.
\item We report the oracle's behavior in isolation of the cutting-plane loop under finite sampling, under a device noise model, and on quantum hardware as well as a full execution of the loop under the noise model. To our knowledge this is the first study to embed a variational eigensolver within a cutting-plane framework for SDP and to report computational evidence across this range of execution conditions.
\end{itemize}

The remainder of the paper is organized as follows. Section~\ref{sec:literature} reviews related work and positions our contribution within it. Section~\ref{sec:methodology} presents the cutting-plane framework, the VQE background it relies on, and the three components specific to our oracle. Section~\ref{sec:computational-study} reports the computational study, and Section~\ref{sec:discussion} discusses the implications for the space-complexity motivation and outlines directions for future work.

\section{Related literature} \label{sec:literature}
The prospect of quantum advantage has prompted renewed analysis of classical optimization algorithms. Many of these quantum approaches remain untested due to current hardware limitations or lack thereof (e.g. unavailability of QRAMs), and they focus on addressing how such a quantum advantage may be realized in the future. Majority of the existing work in continuous optimization remains to be theoretical. For linear programming (LP), interior-point methods~\cite{mohammadisiahroudi2022efficient,apers2023quantum,augustino2023quantum}, dual logarithmic barrier~\cite{wu2025quantumIJOO} and the simplex method~\cite{nannicini2024fastLP} are adapted. 
Interior-point methods are also developed for second-order cone programs~\cite{augustino2021inexactsocp}, linearly constrained quadratic programs~\cite{wu2023inexact} and SDPs~\cite{brandao2017quantumspeedup,van2017quantum,brandao2017quantum,van2018improvements,kerenidis2020quantum,augustino2023quantumSDP,mohammadisiahroudi2025quantum}.

The (theoretical) quantum advantage in many of these papers stems from quantizing linear algebra operations, in particular, solving linear systems using quantum linear algebra routines. However, for the most part, these methodologies have not been tested on quantum hardware or simulated on classical computers. 

Recent work demonstrates the potential of hybrid quantum-classical optimization, embedding quantum or Ising solvers within classical branch-and-bound, branch-and-price, and cutting-plane frameworks for problems such as quadratic unconstrained binary optimization (QUBO), traveling salesperson problem (TSP), vertex coloring, electric vehicle charging, and power-system optimization~\cite{peng2025hybridquantumbranchandboundmethod,ciacco2026cuttingplanemethodologyquantumoptimization,vercellino2025hybridquantumclassicalbranchandpricemethod,sun2026hybridquantumclassicalbranchandpriceintraday, ellinas2024hybridquantumclassicalalgorithmmixedinteger}. These studies suggest that quantum subroutines can address difficult combinatorial components while classical methods retain the overall optimization structure; together with theoretical foundations such as the copositive framework of~\cite{bernal2024copositivehybrid}, they provide encouraging evidence for integrating quantum optimization into classical exact algorithms.

Research presenting computational evidence of quantum algorithms for SDP problems is quite limited. Early work on quantum SDP uses quantum matrix multiplicative weights techniques, with thermal pure quantum states proposed to reduce the cost of Gibbs-state preparation~\cite{watts2023quantumsdp}. More recent approaches adopt variational quantum algorithms, reformulating SDP constraints through slack variables or Lagrangian penalties and solving the resulting problems with parameterized quantum circuits~\cite{chen2025slackSDP}. Related work extends VQE-based constrained optimization to sparse conic programs, including optimal power flow, exploiting graph structure to reduce quantum resources~\cite{Le2024VQEconstrained,le2026solvingconicprogramssparse}. In parallel, quantum thermodynamics has provided an alternative formulation of SDP through free-energy minimization~\cite{liu2025quantumthermodynamicssemidefiniteoptimization}. Most recently, the authors of~\cite{Nie2026quantumalternating} develop a quantum ADMM framework for SDP, combining classical ADMM with quantum singular-value transformation to accelerate key linear-algebraic operations.
Two lines of existing work are closest to ours: (1) the randomized cutting-plane method of~\cite{marecek2021cutting}, which invokes a quantum eigensolver inside a boundary oracle, and (2) the inexact variational algorithm of~\cite{patel2024variational}, which reformulates the SDP itself as a saddle-point problem solved approximately by quantum circuits. Neither is a cutting-plane algorithm of the type proposed here.

To the best of our knowledge, ours is the first study to use VQE as a separation oracle in an SDP cutting-plane scheme and to report computational evidence spanning noiseless simulation, noisy simulation and hardware execution of the oracle. The framework is also general enough to be adapted to other problem classes, including mixed-integer SDPs.

The closest prior work to ours is~\cite{marecek2021cutting}, which proposes a randomized cutting-plane (RCP) method for SDP built on hit-and-run sampling in the spirit of Bertsimas and Vempala~\cite{bertvempala2004_convexprograms_randomwalks}, with a quantum eigensolver invoked inside a boundary oracle that solves a generalized eigenvalue problem to locate the exit point of a sampled ray from the feasible spectrahedron. Their quantum component is not implemented: they emulate an imperfect quantum eigensolver by injecting synthetic multiplicative and additive noise into classically computed eigenvalues, and study the resulting robustness of RCP to oracle error. Our work differs along several axes. Algorithmically, we adopt a polyhedral outer-approximation framework rather than a randomized sampling scheme, and our oracle directly certifies (non-)PSDness of the candidate returned by the relaxation rather than solving a generalized eigen-problem along a sampled direction. Most importantly, we implement and run an actual VQE as the separation oracle and report its estimates against the exact minimum eigenvalue at every cutting-plane iteration, which to our knowledge is the first such empirical characterization in this setting. Because the exact eigendecomposition oracle is run within an identical outer loop, the effect of oracle choice is isolated from that of relaxation design. Related but distinct is the variational-quantum-algorithm line of~\cite{patel2024variational} and its extensions~\cite{Le2024VQEconstrained,le2026solvingconicprogramssparse, Westerheim2026DualVQE}, which reformulate the SDP itself as a variational optimization problem over a fixed ansatz rather than using a quantum subroutine inside a classical cutting-plane loop, and which target a different SDP standard form prevalent in quantum information applications.

\section{A cutting-plane framework with a variational quantum separation oracle} \label{sec:methodology}
We solve the standard-form SDP \eqref{eq:sdp-generic} by a cutting-plane scheme that relaxes the conic constraint $X \in \mathbb{S}^n_+$ to the symmetry requirement $X \in \mathbb{S}^n$ and recovers PSDness through valid inequalities generated dynamically. Each iteration solves a tractable relaxation, which is a linear or a second-order cone program, to obtain a candidate $X^*$, then calls a separation oracle that either certifies $X^* \succeq 0$ or returns a direction of violation from which a cut is constructed. The distinguishing feature of our proposed cutting plane approach is that the separation oracle is realized by VQE, which estimates the minimum eigenvalue of $X^*$, denoted by $\lambda_{\min}(X^*)$, using a number of qubits logarithmic in the matrix dimension. Algorithm~\ref{alg:generic-cutting-plane} states the scheme in a form that is independent of the choice of initial outer approximation, separation oracle, and cut selection rule, which are specified in the subsections below.

\begin{algorithm}[htbp]
\caption{Generic cutting-plane scheme for the SDP \eqref{eq:sdp-generic}.}
\label{alg:generic-cutting-plane}
\begin{algorithmic}[1]
\REQUIRE Initial outer approximation $\mathcal{C}_0 \supseteq \mathbb{S}^n_+$
         (Section~\ref{sec:methodology_relaxation});
         separation oracle $\mathcal{O}$ (Section~\ref{sec:methodology_separation});
         cut selection rule \textsc{SelectCuts} (Section~\ref{sec:cut_generation});
         tolerance $\varepsilon > 0$; iteration limit $T$.
\ENSURE Candidate solution $X^*$ and termination status.
\STATE $\mathcal{C} \gets \mathcal{C}_0$
\FOR{$t = 1, \dots, T$}
    \STATE Solve the master problem
           $\sup \{\, A_0 \bullet X : A_k \bullet X = b_k,\ k = 1,\dots,m,\ X \in \mathcal{C} \,\}$.
    \IF{the master problem is infeasible}
        \RETURN \textsc{Infeasible} \COMMENT{the SDP \eqref{eq:sdp-generic} is infeasible}
    \ELSIF{the master problem is unbounded}
        \STATE Let $X^*$ be a ray of unboundedness of the master problem.
    \ELSE
        \STATE Let $X^*$ be an optimal solution of the master problem.
    \ENDIF
    \STATE $(\hat{\lambda}, \mathcal{U}) \gets \mathcal{O}(X^*)$
           \COMMENT{estimate of $\lambda_{\min}(X^*)$ and violating directions}
    \IF{$\hat{\lambda} \ge -\varepsilon$}
        \IF{the master problem was unbounded}
            \RETURN \textsc{Unbounded}
               \COMMENT{$X^*$ is a PSD improving ray; \eqref{eq:sdp-generic} is unbounded}
        \ELSE
            \RETURN $X^*$, \textsc{Optimal}
               \COMMENT{$X^*$ is feasible, hence optimal, for \eqref{eq:sdp-generic}}
        \ENDIF
    \ENDIF
    \STATE $\mathcal{V} \gets \textsc{SelectCuts}(\mathcal{U}, \mathcal{C})$
       \COMMENT{cut family and redundancy control; Section~\ref{sec:cut_generation}}
    \STATE $\mathcal{C} \gets \mathcal{C} \cap \bigcap_{V \in \mathcal{V}} V$
\ENDFOR
\RETURN $X^*$, \textsc{IterationLimit}
\end{algorithmic}
\end{algorithm}
Note that handling the unbounded case requires the master solver to return an improving ray. Not all solvers expose one for conic models, which constrains the choice of solver and of initial outer approximation; we return to this point in Section~\ref{sec:computational-study}.

Algorithm~\ref{alg:generic-cutting-plane} does not have a finite-convergence guarantee. Each generated cut separates the current iterate, but $\mathbb{S}^n_+$ is not polyhedral and cannot be described by finitely many linear or second-order conic inequalities, so in general the scheme approaches an optimal solution of \eqref{eq:sdp-generic} only asymptotically. Termination in \textsc{Optimal} status occurs when a candidate is certified to be PSD  within the tolerance $\varepsilon$; otherwise, the iteration limit $T$ governs, and the objective value of the final master problem provides a valid upper bound on the optimal value of \eqref{eq:sdp-generic}. Infeasibility of the master problem, by contrast, certifies infeasibility of \eqref{eq:sdp-generic}, since $\mathcal{C} \supseteq \mathbb{S}^n_+$ at every iteration.

The correctness of the \textsc{Optimal} status depends on the exactness of the separation oracle $\mathcal{O}$. When  $\mathcal{O}$ returns $\lambda_{\min}(X^*)$ exactly, the test $\hat \lambda \geq -\varepsilon$ certifies that $X^*$ is feasible up to tolerance, and the returned solution is optimal for~\eqref{eq:sdp-generic}. For an inexact oracle this is no longer guaranteed, since for the ones we consider, the estimate satisfies $\hat\lambda \geq \lambda_{\min}(X^*)$ and the test may be passed while $\lambda_{\min}(X^*) < -\varepsilon$, terminating the scheme at an infeasible iterate. We refer to such an outcome as a \emph{false PSDness certificate}. Since the estimate can only overestimate, the error is one-sided: an inexact oracle may stop  early, but never rejects a genuinely PSD iterate. In our computational study every \textsc{Optimal} termination obtained with an inexact oracle is verified a posteriori against an exact eigendecomposition, and false certificates are reported separately in Section~\ref{sec:computational-study}.

\subsection{Relaxation and initial outer approximations} \label{sec:methodology_relaxation}

We  initialize the master problem with a polyhedral or a second-order conic outer approximation of the PSD cone $\mathbb{S}^n_+$ . 
In particular, we enforce the diagonal nonnegativity constraints 
 $   X_{ii} \geq 0 $ 
 $i=1,\dots,n$, 
and add one of the two families of valid inequalities below, derived from the $2\times2$ principal submatrices:
\begin{itemize}
    \item \textbf{Linear family:} $\begin{bmatrix}1\\a\end{bmatrix}^{\!\top} \begin{bmatrix}X_{ii}&X_{ij}\\X_{ij}&X_{jj}\end{bmatrix}\begin{bmatrix}1\\a\end{bmatrix} \ge 0$, for   $1 \le i < j \le n$ and $a \in \mathcal{A}_K$ where $\mathcal{A}_K:=\left\{ \tan\left(\frac{k\pi}{2K}\right) : k = \pm 1, \pm 2, \dots, \pm (K-1)\right\}$. 
    Following \cite{wang2021polyhedral}, we take $K=4$, which yields 
    $\mathcal{A}_4=\{\pm 1, \pm (\sqrt2+1), \pm(\sqrt2-1)\}$.
    \item \textbf{SOC family:} $\begin{bmatrix}X_{ii}&X_{ij}\\X_{ij}&X_{jj}\end{bmatrix} \succeq 0$, for each $1 \le i < j \le n$, which is the exact conic condition on the $2\times2$ principal submatrix. 
\end{itemize}

The linear family enforces the $2\times 2$ conic condition along finitely many directions $(1,a)^\top$, and is therefore implied by the SOC family, which enforces the condition for all directions. Consequently, we consider two initial outer approximations obtained by introducing either one of the two families; in Algorithm~\ref{alg:generic-cutting-plane}, $\mathcal{C}_0$ is the intersection of $\mathbb{S}^n$ with the diagonal constraints and one of the two families above.

\subsection{The separation problem and oracles} \label{sec:methodology_separation}

Given a candidate solution $X^* \in \mathbb{S}^n$, the separation oracle must determine whether $X^* \succeq 0$ and, if not, return a vector~$u$ with $u^\top X^* u < 0$. Such a vector~$u$ yields the linear inequality
  $  u^\top X u \ge 0$,
which is satisfied by every $X \succeq 0$ but violated by the current $X^*$, and hence is a valid cut. A canonical choice for $u$ is an eigenvector of $X^*$ corresponding to a negative eigenvalue (if one exists); in particular, an eigenvector associated with the most negative eigenvalue produces the most violated cut of this form among all unit vectors. The matrix is certified to be PSD, i.e., $X^* \succeq 0$ if and only if $\lambda_{\min}(X^*) \ge 0$. 

When the matrix variable is block-diagonal, the separation problem decomposes: $X^* \succeq 0$ if and only if each diagonal block is PSD. The oracle is therefore invoked once per block, and a violating direction found for a given block yields a cut on that block's entries alone. In Algorithm~\ref{alg:generic-cutting-plane}, the separation step of line~10 and the cut generation of line~16 are performed for each block in turn, and the scheme terminates in \textsc{Optimal} status only when every block passes the PSDness test of line~11. In the remainder of the paper, $n$ denotes the dimension of the block under consideration. 

We consider the following four realizations of the separation oracle:
\begin{enumerate}
    \item \textbf{Eigenvalue decomposition (EVD):} Computes the eigenspectrum of $X^*$ exactly, yielding all eigenpairs at a cost of $O(n^3)$ arithmetic operations, and used as the reference oracle throughout. 
    \item \textbf{Quadratic programming (QP):} Solves the (potentially nonconvex) quadratic optimization problem given by 
\begin{equation}\label{eq:quad_prog}
            z^* = \min\{u^\top X^* u : \|u\|_2 \le 1\},
\end{equation}
    which certifies that $X^* \succeq 0$ when $z^* = 0$; otherwise an optimal solution $u^*$ satisfies $u^{*\top}X^* u^*<0$ and yields a cut. 
    \item \textbf{QUBO approximation:} Discretizes the quadratic program~\eqref{eq:quad_prog} using a binary-encoding approach to approximate $u$ to $H$ bits of precision, yielding a quadratic unconstrained binary optimization (QUBO) problem with $nH$ binary variables. As before, a negative optimal value certifies that $X^*$ is not PSD; however, the test may be inconclusive for small $H$.
    \item \textbf{Variational quantum eigensolver (VQE):} Estimates $\lambda_{\min}(X^*)$ using a variational quantum approach, which is applicable since $X^*$ is real symmetric and hence Hermitian.
\end{enumerate}

Although the QP- and QUBO-based oracles above are conceptually valid approaches to solve the separation problem, preliminary analysis showed both to be impractical within the cutting-plane loop. Solving the nonconvex QP to global optimality, and solving its QUBO approximation for a range of precision parameters $H$, incurred separation times that are prohibitively large, and the resulting improvement in the master problem's objective value across iterations was correspondingly slow. We therefore exclude the QP- and QUBO-based oracles from the computational study of Section~\ref{sec:computational-study}, and focus our analysis on the EVD and VQE oracles. Background information on the VQE approach in general as well as specific details regarding our VQE oracle follow in Sections~\ref{sec:methodology_vqe} and \ref{sec:methodology_vqe_oracle}.

\subsection{Variational quantum eigensolver} \label{sec:methodology_vqe}

We now describe VQE, 
the algorithm underlying our quantum separation oracle. Our exposition emphasizes the linear-algebraic content of the method; we refer the reader to \cite{cerezo2021variational} for a broader survey of variational quantum algorithms.

\subsubsection{States, observables, and the variational principle} \label{sec:quantum_basics}

A quantum register of $q$ qubits is described by a \emph{state vector} $\ket{\psi} \in \mathbb{C}^{N}$ with $N = 2^{q}$ and $\|\ket{\psi}\|_2 = 1$; its coordinates in the \emph{computational basis} are its \emph{amplitudes}. An \emph{observable} is a Hermitian operator $\hat{H} \in \mathbb{C}^{N \times N}$, and its \emph{expectation value} in the state $\ket{\psi}$ is the quadratic form
\begin{equation} \label{eq:quad_expectation}
    \braket{\hat{H}}_{\ket{\psi}}:= \bra{\psi} \hat{H} \ket{\psi}
    = \sum_{k=0}^{N-1} \lambda_k |\alpha_k|^2 ,
\end{equation}
where $\lambda_0 \le \cdots \le \lambda_{N-1}$ are the eigenvalues of $\hat H$ and $\alpha_k$ the coefficients of $\ket{\psi}$ in the corresponding eigenbasis: measuring $\hat H$ in the state $\ket\psi$ returns $\lambda_k$ with probability $|\alpha_k|^2$. For real symmetric $\hat H$ and real amplitudes $\tilde u \in \mathbb{R}^N$ with $\|\tilde u\|_2 = 1$, \eqref{eq:quad_expectation} is the Rayleigh quotient $\tilde{u}^{\top} \hat{H} \tilde{u}$, whose minimum over the unit sphere is $\lambda_{\min}(\hat H)$. Since the $|\alpha_k|^2$ sum to one and $\lambda_0 \le \lambda_k$ for every $k$, \eqref{eq:quad_expectation} yields the \emph{variational principle}: $\braket{\hat{H}}_{\vec{\theta}} := \bra{\psi(\vec{\theta})} \hat{H} \ket{\psi(\vec{\theta})} \ge \lambda_0$ for any normalized parameterized state and all $\vec{\theta} \in \Theta$, with equality if and only if the family contains a ground state of $\hat H$.

The relationship to the classical oracles of Section~\ref{sec:methodology_separation} is direct: the QP-based oracle minimizes the Rayleigh quotient over the entire unit ball, whereas VQE minimizes it over the subset of the unit sphere reachable by a parameterized quantum circuit, called an \emph{ansatz}. This restriction makes the problem tractable on quantum hardware, and it is also the source of the method's inexactness: whenever the reachable set does not contain a ground state of $X^*$, the returned estimate is a strict upper bound on $\lambda_{\min}(X^*)$. The estimate is inexact even when it does, unless the minimization of the expectation value is itself exact.

\subsubsection{Algorithmic components}\label{sec:algo_components}

Being a variational quantum algorithm, the VQE is a hybrid approach where a quantum circuit is iteratively trained to prepare a state that encodes a solution to the problem of interest. It is built from three components \cite{cerezo2021variational}:

\begin{enumerate}
    \item \textbf{Ansatz.} A parameterized quantum circuit $U_A(\vec{\theta})$ defining the search space, i.e.\ the family of trial states $\ket{\psi(\vec{\theta})} = U_A(\vec{\theta}) \ket{0}$ accessible by varying $\vec{\theta}$. We often write $U_A(\vec{\theta}) = U_V(\vec{\theta}) U_R$, where a non-parameterized reference operator $U_R$ prepares a fixed reference state and the variational form $U_V(\vec{\theta})$ carries the free parameters.

    \item \textbf{Estimator.} A routine that evaluates the \emph{cost function}
    $
        f(\vec{\theta}) = \bra{\psi(\vec{\theta})} \hat{H} \ket{\psi(\vec{\theta})}
    $.
    Because a quantum measurement returns a sample (bitstring) rather than this expectation value, the circuit is executed repeatedly---each execution is called a \emph{shot}---and the outcomes are then combined classically to estimate $f(\vec\theta)$. The decomposition of $\hat H$ that makes this possible is described in Section~\ref{sec:cost_func_eval}.
    
    \item \textbf{Classical optimizer.} A  routine that solves the problem $\min_{\vec{\theta} \in \Theta} f(\vec{\theta})$ classically using the estimator's output, where the parameter set $\Theta$ is taken to be $[0,2\pi]^{n_\theta}$ and $n_\theta$ is the number of variational parameters, the angles being $2\pi$-periodic. Both derivative-free and gradient-based methods are commonly used.
\end{enumerate}

These components interact in a hybrid loop: the quantum processor samples measurement outcomes from $\ket{\psi(\vec{\theta})}$, and the classical processor estimates $f(\vec{\theta})$ and updates $\vec\theta$ until the optimizer's stopping criterion is met. At termination, the state $\ket{\psi(\vec{\hat\theta})}$ and the value $f(\vec{\hat\theta})$ approximate a ground state of $\hat H$ and its minimum eigenvalue $\lambda_0$.

The readout step is more demanding here than in a typical VQE application, where the eigenvalue or samples from the optimized circuit suffice. Forming a cut requires the \emph{amplitudes} of the optimized state, which on hardware would call for \emph{state tomography}, whose cost is exponential in the number of qubits. Whether a cheaper readout scheme sufficient for cut generation exists is an open question, to which we return in Section~\ref{sec:discussion}.

\subsubsection{Cost function evaluation}\label{sec:cost_func_eval}

To evaluate \eqref{eq:quad_expectation} on hardware, the observable is expressed in the Pauli basis
\begin{equation} \label{eq:pauli}
    \hat{H} = \sum_{i} c_i P_i ,
    \qquad
    P_i \in \{ I, X, Y, Z \}^{\otimes q}, \; c_i \in \mathbb{R},
\end{equation}
which forms an orthogonal basis of the real vector space of Hermitian $N \times N$ matrices. By linearity of the expectation, we have
$
    \braket{\hat{H}}_{\vec\theta} = \sum_i c_i \braket{P_i}_{\vec\theta}
$, 
so each term is estimated separately and the results are combined. The number of terms in \eqref{eq:pauli} is at most $N^2 = 4^{q}$ in general. It is bounded above by $N(N+1)/2$ in our case as our observables are always represented by real-valued matrices and hence expressible using only the Pauli strings containing an even number of $Y$ factors. The actual number depends on the (sparsity) structure of $\hat{H}$; we report the Pauli term counts arising in our instances in Section~\ref{sec:encoding-cost}.

Two features of this estimation procedure are relevant to the behavior we observe in Section~\ref{sec:computational-study}. First, the estimate of each $\braket{P_i}_{\vec\theta}$ is a sample mean, so its statistical error decreases as $O(1/\sqrt{S})$ in the number of shots $S$; attaining an accuracy $\delta$ therefore requires $O(1/\delta^2)$ circuit executions. Consequently, the accuracy with which VQE can resolve a small $|\lambda_{\min}(X^*)|$ is limited by the available measurement budget, in contrast to the classical EVD oracle, which returns the eigenvalue to machine precision. Second, each cost function value requested by the classical optimizer requires one call to the estimator, and the optimizer is invoked once per separation call; the number of circuit executions per cutting-plane iteration therefore scales with the product of the shot count and the number of function evaluations, up to the number of distinct measurement circuits required for $\hat H$.

\subsubsection{Ansatz design and expressivity}\label{sec:ansatz_design}

The efficacy of VQE is governed by the ansatz design, which determines the reachable subset of state space and hence the quality of the attainable bound. A more expressive ansatz covers a larger portion of the state space and is more likely to contain a good approximation of the ground state, but does so at the cost of additional parameters and circuit depth.

The standard choice is the \emph{hardware-efficient} ansatz (HEA) \cite{kandala2017hardware}, a problem-independent circuit of single-qubit rotation layers followed by a fixed (e.g.\ linear) pattern of two-qubit entangling gates, chosen for ease of execution on near-term devices. Since it incorporates no information about $\hat{H}$, the classical optimizer must search a high-dimensional landscape without structural guidance. Increasing the number of layers to improve expressivity aggravates a second difficulty: for sufficiently expressive randomly initialized circuits, the variance of the cost gradient decays exponentially in the number of qubits, a phenomenon known as the \emph{barren plateau} \cite{mcclean2018barren}, in which the landscape is effectively flat. These considerations motivate problem-inspired ans\"atze that reflect the structure of the operator; the construction we introduce in Section~\ref{sec:ansatz} derives the entanglement pattern from the candidate matrix rather than fixing it in advance.

\subsubsection{Resource scaling}\label{sec:res_scaling}

The property that motivates the use of VQE as a separation oracle is its qubit requirement. Estimating the extremal eigenpair of an $N \times N$ Hermitian matrix requires a register of $q = \log_2 N$ qubits, since a $q$-qubit state vector has $2^{q}$ amplitudes. A matrix of order $n$ is therefore handled with $\lceil \log_2 n \rceil$ qubits, in contrast to the $O(n^2)$ storage required to represent the matrix classically. We emphasize that this logarithmic scaling concerns the qubit count alone; the costs of preparing the operator in the form \eqref{eq:pauli}, of executing sufficiently many shots, and of extracting the amplitudes of the final state $\ket{\psi(\vec{\hat\theta})}$ are separate questions, which we take up in Sections~\ref{sec:encoding-cost} and~\ref{sec:discussion}.

\subsection{Proposed VQE oracle} \label{sec:methodology_vqe_oracle}
Having discussed the fundamentals of the VQE, we now describe the non-standard features of the proposed VQE oracle embedded within our cutting plane framework. 

\subsubsection{Zero-padding} \label{sec:padding}

The blocks of the candidate matrix have dimensions determined by the SDP instance and are not in general powers of two. We therefore embed $X^*$ into the next power-of-two dimension $N = 2^{\lceil \log_2 n\rceil}$ by zero-padding:
\begin{equation}\label{eq:zero-padding}
    X^\prime = \begin{bmatrix} X^* & 0 \\ 0 & 0 \end{bmatrix} \in \mathbb{S}^N.
\end{equation}
VQE operates on $X^\prime$ and returns a violating vector $\tilde u \in \mathbb{R}^N$, which we partition as $\tilde u = (u, w)$ with $u \in \mathbb{R}^n$ and $w \in \mathbb{R}^{N-n}$. The cut is formed from the truncated vector~$u$.
Note that this truncation is exact since
we have
\begin{equation}\label{eq:truncation}
   \tilde u^\top X^\prime \tilde u = u^\top X^* u
\end{equation}
for every $\tilde u$, independently of $w$. Consequently, whenever VQE reports $\tilde u^\top X^\prime\tilde u < 0$, the truncated vector satisfies $u^\top X^* u < 0$ exactly and thus $u$ yields a valid violated cut $u^\top X u \ge 0$ for the original problem. 

When $N>n$, the padding introduces an eigenspace of dimension $N-n$ at eigenvalue $0$. This does not affect correctness, but it introduces a subspace on which the quadratic form vanishes identically: a trial state placing most of its probability mass there returns an estimate near zero irrespective of $\lambda_{\min}(X^*)$. We quantify this effect in Section~\ref{sec:main-results}, where it proves to be the dominant failure mode of the oracle on the larger instances.

\subsubsection{Operator-derived ansatz} \label{sec:ansatz}

As mentioned earlier, the HEA applies single-qubit rotations followed by a fixed (e.g. linear) chain of entangling gates, independent of the operator whose eigenvalue is to be estimated. This ignores structure in the candidate matrix that could guide where entanglement is most useful. We therefore introduce an \emph{operator-derived} ansatz (ODA), whose entangling connectivity is induced by the operator itself. Specifically, we take the entanglement map from the qubit supports of the Pauli decomposition \eqref{eq:pauli}: an entangling gate is placed between qubits $i$ and $j$ whenever some term $P_k$ with $c_k \neq 0$ acts non-trivially on both qubits.

The resulting connectivity therefore adapts to whatever structure the candidate matrix exhibits: a structured operator yields a sparse entanglement map, while one whose Pauli support involves every qubit pair yields the complete graph and hence full entanglement. Whether the maps arising in practice are sparse or complete is examined empirically in Section~\ref{sec:encoding-cost}. Since the map need not remain fixed across iterations, we cache circuits by entanglement map, rebuilding only when a new map is encountered to avoid unnecessary transpilation overhead.

A further consideration specific to our setting is that the observable is always real symmetric, since $X^* \in \mathbb{S}^n$ and padding preserves symmetry, so $X^\prime$ admits an orthonormal basis of \emph{real} eigenvectors. Restricting the trial states to real amplitudes therefore entails no loss: the target state remains inside the restricted search space.

We therefore employ ans\"atze composed of $R_y$ rotations and $\mathrm{CX}$ entanglers acting on $\ket{0}$, that is $U_R=I$, which generate exactly the real-amplitude states (the \texttt{RealAmplitudes} family in the Qiskit circuit library). Besides guaranteeing that the violating directions are real-valued and hence usable without further processing, this halves the parameterization: with $q$ qubits and $p$ entangling layers the circuit carries $q(p+1)$ parameters, against $2q(p+1)$ for a complex-capable ansatz such as \texttt{EfficientSU2}, whose additional $R_z$ rotations introduce only relative phases. Since the parameter count depends on $q$ and $p$ alone, ODA and HEA are matched in parameters and differ only in their entangling topology.

\subsubsection{Cut generation strategy} \label{sec:cut_generation}

Each violating direction returned by the oracle yields a linear cut. When SOC cut generation is enabled, we additionally form a single second-order conic cut per block per iteration, provided at least two violating directions are available. Let $\tilde u_1, \tilde u_2$ be the two directions with the most negative values of $\tilde u^\top X^\prime \tilde u$, and $u_1, u_2$ their truncations to $\mathbb{R}^n$. The resulting cut is
\begin{equation} \label{eq:soccut}
    \begin{bmatrix} u_1^\top \\ u_2^\top \end{bmatrix}
    X
    \begin{bmatrix} u_1 & u_2 \end{bmatrix} \succeq 0,
\end{equation}
which is valid for all $X \succeq 0$. Since every violating direction found in a single separation call yields a cut at no additional oracle cost, it is advantageous to extract as many as possible per call.

For the EVD oracle this is immediate: the eigendecomposition of $X^*$ exposes its entire eigenspectrum, so every eigenvector associated with a negative eigenvalue is a violating direction that produces a linear cut, and a second-order conic cut \eqref{eq:soccut} may be formed from the eigenvectors associated with the two most negative eigenvalues. 

For the VQE oracle it is less obvious, since the algorithm is designed to return a single approximate ground state. Nevertheless, multiple violating directions can be obtained from a single call without any additional cost function evaluations, as follows. Recall that the cost function minimized by the classical optimizer is the expectation value $f(\vec{\theta}) = \bra{\psi(\vec{\theta})} X^\prime \ket{\psi(\vec{\theta})}$. By construction, \emph{any} parameter vector $\vec{\theta}$ with $f(\vec{\theta}) < 0$ certifies that the associated trial state $\ket{\psi(\vec{\theta})}$ provides a direction of violation, even if it is not the optimal one. We therefore attach a callback to the classical optimizer which records every iterate $\vec{\theta}_t$ at which $f(\vec{\theta}_t) < 0$. Binding the ansatz to each such $\vec{\theta}_t$ and reading out the resulting state yields a collection of violating vectors 
 $   \widetilde{\mathcal{U}} = \big\{\, \tilde u(\vec{\theta}_t) \;:\; f(\vec{\theta}_t) < 0 \,\big\} \subset \mathbb{R}^N,
 $
where $\tilde u(\vec\theta)$ denotes the amplitude vector of $\ket{\psi(\vec\theta)}$ in the computational basis, which is real by the construction of Section~\ref{sec:ansatz}. Each element satisfies
$\tilde u^\top X^\prime \tilde u < 0$, thereby producing a valid cut after truncation to the original dimension as described in Section~\ref{sec:padding}; it is the set of truncated vectors that is returned by the oracle as $\mathcal{U}$ in Algorithm~\ref{alg:generic-cutting-plane}. The expectation values $f(\vec{\theta}_t)$ are computed in the course of the optimization regardless, so no further circuit executions are needed; the only additional cost is the readout of the corresponding states, which is inexpensive in simulation but, as noted in Section~\ref{sec:algo_components}, demanding on hardware.

Note that the iterates $\vec{\theta}_t$ collected along a single optimizer trajectory are correlated by construction: as the optimizer converges, successive trial states approach a common limit, and the associated cuts become increasingly parallel. Because near-parallel cuts contribute little while enlarging the master problem, each candidate cut is compared against the previously generated ones by cosine similarity and discarded when its overlap with an existing cut exceeds a threshold $\tau \in (0,1)$. Multi-cut collection from a variational oracle requires such filtering to a greater extent than the EVD oracle does, whose eigenvector cuts are mutually orthogonal within a single cutting plane iteration, so that filtering is relevant only across iterations. 

We emphasize that this filter governs only whether a cut is \emph{added} to the master problem; it plays no role in the termination decision. A block is certified PSD only when the call produces no violating direction at all: neither among the iterates collected along the trajectory, nor at the optimizer's final iterate, where the test $\hat\lambda \ge -\varepsilon$ in Algorithm~\ref{alg:generic-cutting-plane} applies.
The two conditions are distinct: the oracle may detect a violation, $\hat\lambda < -\varepsilon$, and yet every candidate direction may significantly overlap a cut already present in the master problem. Treating such an iteration as a certificate of PSDness would be incorrect, since the current iterate is demonstrably infeasible. We therefore override the filter and admit the direction associated with the optimizer's final iterate regardless of its overlap with existing cuts, so that a detected violation always yields at least one cut.

\section{Computational study} \label{sec:computational-study}
This section reports our computational experience with the framework of Section~\ref{sec:methodology}. We first fix the master problem and VQE oracle configurations through a sequence of controlled experiments, then compare the classical and variational oracles across the instance family, and finally examine the effect of finite sampling and device noise.

\subsection{Experimental setup}\label{sec:setup}

\subsubsection{Test instances}

We evaluate the framework on the \texttt{control} family of the SDPLIB benchmark collection~\cite{borchers1999sdplib}, comprising ten instances of increasing size arising from control-theoretic Lyapunov stability problems as reported in Table~\ref{tab:instances}. 

\begin{table}[htbp]\footnotesize
\caption{SDPLIB \texttt{control} instances. Each instance has two diagonal blocks; $m$ denotes the number of equality constraints and $n$ the order of the full matrix variable.}
\label{tab:instances}
\centering
\begin{tabular}{r|r|r|r}
Instance & $m$ & $n$ (block dimensions) & Optimal value \\ \hline
\texttt{control1}  &   21 &  15 \ (10, 5)   & 17.78463 \\
\texttt{control2}  &   66 &  30 \ (20, 10)  &  8.3 \\
\texttt{control3}  &  136 &  45 \ (30, 15)  & 13.63327 \\
\texttt{control4}  &  231 &  60 \ (40, 20)  & 19.79423 \\
\texttt{control5}  &  351 &  75 \ (50, 25)  & 16.8836 \\
\texttt{control6}  &  496 &  90 \ (60, 30)  & 37.3044 \\
\texttt{control7}  &  666 & 105 \ (70, 35)  & 20.6251 \\
\texttt{control8}  &  861 & 120 \ (80, 40)  & 20.286 \\
\texttt{control9}  & 1081 & 135 \ (90, 45)  & 14.6754 \\
\texttt{control10} & 1326 & 150 \ (100, 50) & 38.533 \\ \hline
\end{tabular}
\end{table}

This family was selected for two reasons. Since the oracle is applied blockwise, the relevant size parameter is the block dimension rather than the order $n$; families of many small blocks are therefore not of interest, each block yielding an eigenvalue problem a classical routine resolves at negligible cost. The \texttt{control} instances have two blocks of non-trivial dimension, so the separation problem is substantive at every iteration, and the family provides a clean size-scaling axis. Reference optimal values are those reported with the benchmark collection and were verified with \textsc{Mosek}.

\subsubsection{Computational environment}

All experiments were run on two identical virtual machines with an Intel Xeon w5-2465X processor and 10\,GiB of memory under Ubuntu 22.04 LTS; since the operating system and supporting services hold part of it, the effective limit is roughly 8.4\,GiB. The implementation is in Python 3.11.9, with master problems solved by \textsc{Mosek}~11 and \textsc{Gurobi}~13 and quantum circuits built and simulated with Qiskit~2.2.3, \texttt{qiskit-aer}~0.17.2, \texttt{qiskit-algorithms}~0.4.0 and \texttt{qiskit-ibm-runtime}~0.45.1. Unless stated otherwise, expectation values were computed exactly by state vector simulation; the effect of finite sampling and device noise is examined separately in Section~\ref{sec:noisy}.

\subsubsection{Fixed parameter settings}

Unless stated otherwise, all runs use the following settings. The PSDness tolerance of Algorithm~\ref{alg:generic-cutting-plane} is $\varepsilon = 10^{-6}$, and the conic interior-point tolerances of \textsc{Mosek} are likewise set to $10^{-6}$; the motivation for the latter is given in Section~\ref{sec:master-config}. Multi-cut collection is enabled throughout, and the cosine-similarity threshold of the diversity filter is $\tau = 0.9$.

Cut purging is enabled in the scaling study of Section~\ref{sec:main-results} and in the noisy simulations of Section~\ref{sec:noisy}: a linear cut whose dual multiplier remains below $10^{-8}$ for five consecutive iterations is removed from the master problem. Its purpose is mainly to limit the growth of the master problem on larger instances and therefore is disabled in the configuration experiments of Sections~\ref{sec:master-config} and~\ref{sec:vqe-config}.

Before each separation call, entries of the candidate matrix whose magnitude is below $\varepsilon$ are set to zero, removing numerical noise below the tolerance to which the master problem is solved and yielding a sparser operator for the Pauli decomposition \eqref{eq:pauli}. This scrubbed matrix is the one passed to the oracle, so the reported estimates and the entanglement maps are derived from it; the cuts produced remain valid for the original problem, since $u^\top X u \ge 0$ must hold for every $X \succeq 0$ and every $u\in \mathbb{R}^n$.

\subsubsection{Repeated runs}

The EVD oracle and the master solves are deterministic, so runs using the exact oracle are reported from a single execution. The VQE oracle is stochastic --- the variational parameters are initialized at random, and repeated runs follow different optimizer trajectories --- so results involving it are reported as means over repeats, with sample standard deviations where space permits, and the number of repeats is stated in each table. The configuration experiments of Section~\ref{sec:vqe-config} use five repeats on \texttt{control2} and three on \texttt{control5} and \texttt{control6}; the scaling study of Section~\ref{sec:main-results} uses three repeats on \texttt{control1}--\texttt{control5} and a single run on the larger instances, for which the cost of repetition is prohibitive. Since the effects reported for the larger instances are of a different order of magnitude than the run-to-run variation observed on the smaller ones, we do not expect this to affect our conclusions.

\subsection{Master problem configuration}\label{sec:master-config}

The generic scheme of Algorithm~\ref{alg:generic-cutting-plane} leaves open the choices of initial outer approximation, cut family, and the solver used for the master problem. We fix these through an initial experiment on \texttt{control2} using EVD as the separation oracle, so that the comparison isolates the master problem from any effect of oracle
inexactness. All $2^3 = 8$ combinations of solver (\textsc{Gurobi}, \textsc{Mosek}), initial outer approximation (linear, SOC), and second-order cone cut separation (disabled, enabled) were run for $500$ cutting-plane iterations. Results are reported in Table~\ref{tab:master-config}.

\begin{table}[htbp]\footnotesize
\caption{Results on \texttt{control2} with various master problem configurations and EVD. Optimality gap is computed as $(b_f - z^\star)/z^\star$, with $b_f$ and $z^\star$ denoting respectively the final bound and the reference optimal value of \texttt{control2} from Table~\ref{tab:instances}.}
\label{tab:master-config}
\centering
\begin{tabular}{llcrrrrr}
\hline
 &  & SOC & Final & Optimality & Master solve & Separation & \# of cuts  \\
Solver & $\mathcal{C}_0$ & cuts & bound & gap (\%) & time (s) & time (s) & lin./SOC \\
\hline
\textsc{Mosek}  & SOC    & \checkmark & \textbf{88.24} &  963 & \textbf{400.0} & 1.86 & 4099 / 706 \\
\textsc{Gurobi} & SOC    & \checkmark & 89.40 &  977 & 5858.5 & 1.77 & 4098 / 687 \\
\textsc{Gurobi} & linear & \checkmark & 91.29 & 1000 & 7240.2 & 1.74 & 4217 / 714 \\
\textsc{Mosek}  & linear & \checkmark & 93.44 & 1026 &  503.2 & 2.01 & 4185 / 709 \\
\textsc{Gurobi} & SOC    & \ding{55}  & 125.27 & 1409 & 5041.3 & 1.80 & 4010 / --- \\
\textsc{Mosek}  & SOC    & \ding{55}   & 126.16 & 1420 &  192.3 & 1.96 & 3917 / --- \\
\textsc{Mosek}  & linear & \ding{55}   & 127.62 & 1438 &  197.2 & 1.98 & 4122 / --- \\
\textsc{Gurobi} & linear & \ding{55}   & 127.79 & 1440 &  107.4 & 2.05 & 4088 / --- \\
\hline
\end{tabular}
\end{table}

Three observations follow. First, SOC cut separation is by a wide margin the most influential of the three choices: every configuration generating SOC cuts attains a bound roughly $30\%$ better than those restricted to linear cuts, across all combinations of solver and initial relaxation, and the impact grows with the iteration count.

Second, the solver is immaterial for the quality of the bound but decisive for the time required to obtain it: with SOC cut separation enabled, \textsc{Mosek} is roughly 15 times faster than \textsc{Gurobi}, unsurprisingly given that its interior-point method is designed for conic programs. \textsc{Mosek} is also the only one of the two that exposes a ray of unboundedness for conic models, in the form of a dual infeasibility certificate, which the scheme requires when the master problem is unbounded in early iterations.

Third, the initial outer approximation matters least, and its effect is size-dependent: the SOC family yields the better bound on \texttt{control2}, but on \texttt{control5} and \texttt{control6} at $250$ iterations (results not tabulated) the linear family is better by $2.5\%$ and $5\%$, while the SOC family remains $29\%$ to $34\%$ cheaper in master solve time. We therefore adopt \textsc{Mosek} with the SOC initial outer approximation and SOC cut separation enabled throughout, accepting a marginally weaker bound at larger sizes for a substantially cheaper master problem. The conic tolerances of \textsc{Mosek} required adjustment before this configuration could be used at scale: under default settings the master problem became numerically unstable as cuts accumulated and terminated with an unknown status, so we relaxed the primal and dual feasibility tolerances to $10^{-6}$ throughout, the same order as the PSDness tolerance of Algorithm~\ref{alg:generic-cutting-plane}.

In each of the eight configurations, the EVD oracle consumed less than $2\%$ of total runtime across all 500 iterations: separation is effectively free for the classical oracle, and the running time is almost entirely determined by the master problem. This is the baseline against which the cost of the variational oracle is assessed in Section~\ref{sec:main-results}.

Table~\ref{tab:master-config} also shows that the cutting-plane scheme converges slowly even with an exact separation oracle. After $500$ iterations the best configuration attains a bound of $88.24$ against the optimal value of $8.30$, and the master objective is still descending at the iteration limit. The limiting factor is therefore the outer polyhedral scheme rather than the accuracy of the separation oracle.

\subsection{VQE oracle configuration}\label{sec:vqe-config}

With the master problem fixed, we turn to the configuration of the variational oracle: the ansatz family, the number of entangling layers $p$, the classical optimizer, and the optimizer's evaluation budget. Experiments use \texttt{control2} and $250$ cutting-plane iterations with five repeats per configuration unless stated otherwise. Because runs terminating early are not comparable on total separation time, we report separation time per cutting-plane iteration throughout. Results are given in Table~\ref{tab:vqe-config}.

\begin{table}[htbp]\footnotesize
\caption{VQE oracle configurations on \texttt{control2}, $250$ cutting-plane iterations, $\texttt{maxiter}=200$, five repeats. Bounds are means over repeats with sample standard deviations; ``Iter.'' is the mean number of iterations completed before termination; ``FC'' counts runs terminating with a false PSDness certificate.}
\label{tab:vqe-config}
\centering
\begin{tabular}{llcrrrc}
\hline
Ansatz & Optimizer & $p$ & Final bound & Iter. & Sep./iter.\ (s) & FC \\
\hline
HEA & COBYLA & 1 & $231.51 \pm 11.71$ & 250 & 5.61 & 0/5 \\
HEA & COBYLA & 2 & $175.39 \pm 3.81$  & 250 & 5.98 & 0/5 \\
HEA & COBYLA & 3 & $173.61 \pm 3.83$  & 250 & 6.40 & 0/5 \\
ODA & COBYLA & 1 & $229.37 \pm 6.24$  & 250 & 5.79 & 0/5 \\
ODA & COBYLA & 2 & $169.42 \pm 2.12$  & 250 & 6.29 & 0/5 \\
ODA & COBYLA & 3 & $\mathbf{169.17 \pm 2.57}$ & 250 & 6.54 & 0/5 \\
\hline
HEA & SPSA & 2 & $228.80 \pm 9.53$ &  99 & 12.32 & 5/5 \\
HEA & SPSA & 3 & $205.23 \pm 7.96$ & 115 & 12.72 & 5/5 \\
ODA & SPSA & 2 & $237.70 \pm 7.26$ &  89 & 12.64 & 5/5 \\
ODA & SPSA & 3 & $222.24 \pm 8.69$ &  80 & 12.93 & 5/5 \\
\hline
\end{tabular}
\end{table}

The optimizer is selected on reliability rather than bound quality. Every SPSA run terminated prematurely with a false PSDness certificate, whereas COBYLA produced none in $30$ runs. SPSA is also the more expensive of the two per separation call, at roughly twice the cost. Since a separation oracle that frequently certifies an infeasible iterate optimal is not useful within Algorithm~\ref{alg:generic-cutting-plane}, we adopt COBYLA throughout.

With COBYLA the bound improves sharply from $p=1$ to $p=2$ ($229.37$ to $169.42$ for ODA, a $26\%$ reduction) and then saturates: the difference between $p=2$ and $p=3$ is $0.25$ in absolute terms, against run-to-run standard deviations of $2.1$ and $2.6$ respectively. Since this saturation was established only at the block dimensions of \texttt{control2} and the larger instances may require additional expressivity, we fix $p=3$ for the scaling study. Note  that the additional layer costs $4\%$ per separation call~on \texttt{control2}. As the optimizer's evaluation budget is fixed and the per-evaluation cost is mainly determined by the number of Pauli terms, which is independent of $p$, we expect the cost of the additional layer to remain reasonably small at larger block~dimensions. 

At matched optimizer, evaluation budget and $p$ --- hence, by the construction of Section~\ref{sec:ansatz}, at matched parameter count --- the operator-derived ansatz attains a better bound than the hardware-efficient one, by $3.4\%$ at $p=2$ and $2.6\%$ at $p=3$, exceeding the sample standard deviation of either configuration in both cases. Table~\ref{tab:ansatz-scale} repeats the comparison at larger block dimensions. The ordering is preserved, but the margin relative to the run-to-run variation narrows: $1.4\%$ on \texttt{control5} and $2.1\%$ on \texttt{control6}, against comparable standard deviations. Separation cost is indistinguishable between the two ansatz families on every instance tested.

Note here that the comparison reported above is in effect between full and linear entanglement since the candidate matrices were observed to be dense in the Pauli basis throughout the algorithm. In particular, for every instance and block examined, the induced entanglement map was the complete graph on all $q$ qubits, at each iteration. Section~\ref{sec:encoding-cost} reports the Pauli term counts in detail. 

Increasing the number of cost function evaluations per separation call from $200$ to $500$ improves the bound by $5\%$ on \texttt{control2} and $7.4\%$ on \texttt{control6}, at a factor of $2.4$ and $2.5$ respectively in separation cost per iteration (Table~\ref{tab:ansatz-scale}). Since separation dominates the running time, we fix $\texttt{maxiter}=200$ for the scaling study, accepting the looser bound in exchange for tractability on the larger instances.

\begin{table}[htbp]\footnotesize
\caption{Ansatz comparison and effect of the optimizer evaluation budget. COBYLA, $p=3$, $250$ cutting-plane iterations. Five repeats on \texttt{control2} at $\texttt{maxiter}=200$, three otherwise. The $\Delta$ column gives the change in bound from $\texttt{maxiter}=200$ to $500$ for the ODA.}
\label{tab:ansatz-scale}
\centering
\begin{tabular}{llcrrr}
\hline
Instance & Ansatz & \texttt{maxiter} & Final bound & $\Delta$ & Sep./iter.\ (s) \\
\hline
\texttt{control2} & HEA & 200 & $173.61 \pm 3.83$  & ---      &  6.40 \\
\texttt{control2} & ODA & 200 & $169.17 \pm 2.57$  & ---      &  6.54 \\
\texttt{control2} & ODA & 500 & $160.77 \pm 1.36$  & $-5.0\%$ & 15.95 \\
\texttt{control5} & HEA & 200 & $590.03 \pm 3.00$  & ---      & 20.28 \\
\texttt{control5} & ODA & 200 & $581.69 \pm 9.73$  & ---      & 20.09 \\
\texttt{control6} & HEA & 200 & $861.82 \pm 8.93$  & ---      & 19.99 \\
\texttt{control6} & ODA & 200 & $843.92 \pm 11.21$ & ---      & 20.26 \\
\texttt{control6} & ODA & 500 & $781.80 \pm 13.75$ & $-7.4\%$ & 51.66 \\
\hline
\end{tabular}
\end{table}

The configuration adopted for all subsequent experiments is therefore the ODA with $p=3$ entangling layers, optimized by COBYLA with an evaluation budget of $200$ per separation call.

\subsection{Scaling behavior}\label{sec:main-results}

\begin{figure}[htbp]
\centering
\includegraphics[width=0.9\textwidth]{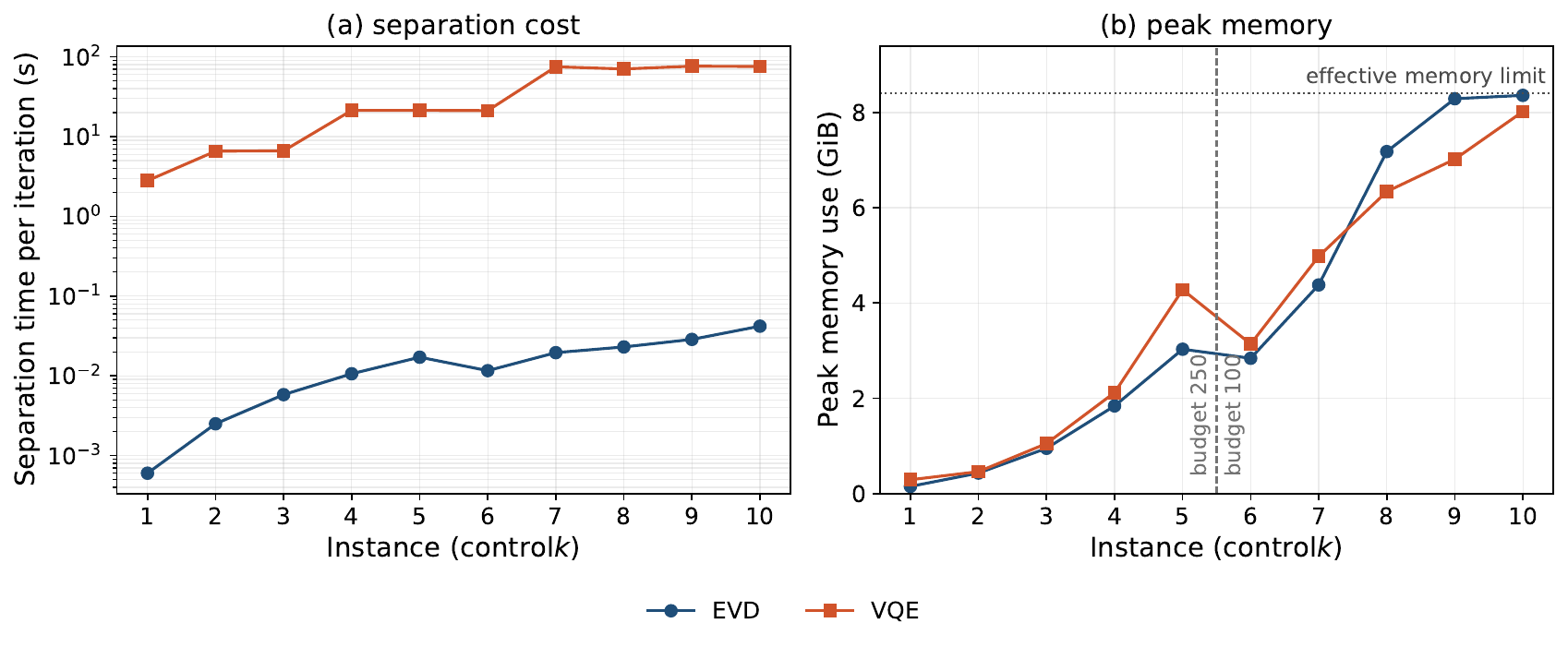}
\caption{Scaling of the two oracles across the instance family. (a) Separation time per cutting-plane iteration, on a logarithmic vertical axis; the plateaus correspond to the qubit count
$q = \lceil \log_2 n\rceil$ of the larger block, which takes the values $4, 5, 6, 7$ across the family. (b) Peak memory use, from the first repeat; the dotted line marks the effective limit at which runs were terminated. The vertical dashed line marks the change in iteration budget, $250$ on \texttt{control1}--\texttt{control5} and $100$ thereafter; peak memory is driven by the accumulated cut pool and is therefore not comparable across it.}
\label{fig:scaling}
\end{figure}

We now compare the two oracles across the family under the configuration fixed in Sections~\ref{sec:master-config} and~\ref{sec:vqe-config}, with iteration budgets of $250$ on \texttt{control1}--\texttt{control5} and $100$ on the larger instances, and three VQE repeats on the smaller instances and one on the larger. On \texttt{control9} and \texttt{control10}, the EVD-based cutting plane solver exhausted the available memory before reaching the iteration budget, terminating at iterations $92$ and $62$ respectively, as did the VQE-based run on \texttt{control10}, at iteration $88$. Results are reported in Table~\ref{tab:scaling} and Figure~\ref{fig:scaling}.

\begin{table}[htbp]\footnotesize
\caption{Scaling comparison of the two oracles. ``Init.'' is the objective value obtained by solving the initial relaxation, identical for both oracles. 
``Closed'' is the fraction of the initial optimality gap closed, $(b_0 - b_f)/(b_0 - z^\star)$, with $b_0$ and $b_f$ denoting respectively the initial and the final bounds, and $z^\star$ the reference optimal value of Table~\ref{tab:instances}. ``Cuts'' gives the respective number of linear and second-order-cone cuts generated, and ``Mem.'' the peak memory use in gigabytes, both from the first repeat for each instance. Bounds for the VQE oracle are means over three repeats on \texttt{control1}--\texttt{control5} --- standard deviations omitted for space --- and single runs thereafter. In cases of memory exhaustion, the number of cutting plane iterations completed is given in parentheses next to the per-iteration separation time; on \texttt{control9} and \texttt{control10} the two oracles stopped at different iterations, so their ``Bound'', ``Cuts'' and ``Closed'' entries are not measured at a common budget.}
\label{tab:scaling}
\centering
\resizebox{\textwidth}{!}{
\begin{tabular}{l| r| rrrrr |rrrrr}
\hline
 & & \multicolumn{5}{c}{EVD} & \multicolumn{5}{|c}{VQE} \\
Instance & Init. & Bound & Sep./iter.\ (s) & Cuts & Mem. & Closed
                 & Bound & Sep./iter.\ (s) & Cuts & Mem. & Closed \\
\hline
\texttt{control1}  &  290.2 &  71.8 & 0.0006 &  678/90  & 0.15 & $80.2\%$
                   &  66.2 &  2.80 &  683/64  & 0.29 & $82.2\%$ \\
\texttt{control2}  &  722.7 & 135.6 & 0.0025 & 2436/389 & 0.43 & $82.2\%$
                   & 169.1 &  6.56 & 1839/229 & 0.46 & $77.5\%$ \\
\texttt{control3}  & 1271.0 & 293.1 & 0.0058 & 4411/499 & 0.95 & $77.8\%$
                   & 354.7 &  6.63 & 2840/304 & 1.05 & $72.9\%$ \\
\texttt{control4}  & 1712.1 & 371.8 & 0.0106 & 6338/500 & 1.84 & $79.2\%$
                   & 487.1 & 21.28 & 5245/369 & 2.12 & $72.4\%$ \\
\texttt{control5}  & 2087.6 & 421.7 & 0.0171 & 7990/500 & 3.03 & $80.4\%$
                   & 584.9 & 21.30 & 6410/361 & 4.28 & $72.6\%$ \\
\texttt{control6}  & 2685.2 & 691.4 & 0.0116 & 4039/200 & 2.84 & $75.3\%$
                   & 1229  & 21.18 & 3436/141 & 3.14 & $55.0\%$ \\
\texttt{control7}  & 3143.8 & 802.2 & 0.0195 & 4725/200 & 4.38 & $75.0\%$
                   & 1703  & 75.08 & 4109/100 & 4.98 & $46.1\%$ \\
\texttt{control8}  & 3540.8 & 902.8 & 0.0230 & 5469/200 & 7.18 & $74.9\%$
                   & 2123  & 70.54 & 4300/100 & 6.34 & $40.3\%$ \\
\texttt{control9}  & 4080.1 & 1037  & 0.0286 \,(92) & 5640/184 & 8.29 & $74.9\%$
                   & 2578  & 76.38 & 4212/100 & 7.02 & $37.0\%$ \\
\texttt{control10} & 4547.3 & 1152  & 0.0419 \,(62) & 4258/124 & 8.36 & $75.3\%$
                   & 3095  & 75.71 \,(88) & 3754/88 & 8.02 & $32.2\%$ \\
\hline
\end{tabular}}
\end{table}

\paragraph{Cost of separation} The variational oracle is three to four orders of magnitude more expensive per separation call than the exact eigendecomposition, the ratio ranging between $1.1 \times 10^3$ and $4.7 \times 10^3$ across the family. Its absolute cost is governed by the qubit count rather than the block dimension (Figure~\ref{fig:scaling}a), consistent with the number of Pauli terms in \eqref{eq:pauli}, which grows as $4^q$ and determines the per-evaluation cost at fixed evaluation budget. Separation consumes under $50$ milliseconds per iteration for the EVD oracle at every size tested, against $76$ seconds for the VQE oracle on the largest instances: the master problem accounts for essentially all of the running time under the exact oracle, whereas under the variational one separation accounts for between $48\%$ and $99\%$ of it.

\paragraph{Quality of the estimates} The bounds obtained with the variational oracle depend on the accuracy with which it resolves the minimum eigenvalue of each block. We therefore compare the returned estimate $\hat\lambda$ against an exact eigendecomposition of the same matrix at every iteration, using the relative error $\left |\hat\lambda - \lambda_{\min}(X^*)\right | / \left |\lambda_{\min}(X^*)\right |$. We also record the \emph{leaked amplitude} $\ell$: the fraction of the optimized state's probability mass lying on the padded coordinates, on which the quadratic form \eqref{eq:truncation} vanishes identically. We write $\epsilon = 1 - \ell$ for its complement, the mass remaining on the original coordinates.

\begin{figure}[htbp]
\centering
\includegraphics[width=0.9\textwidth]{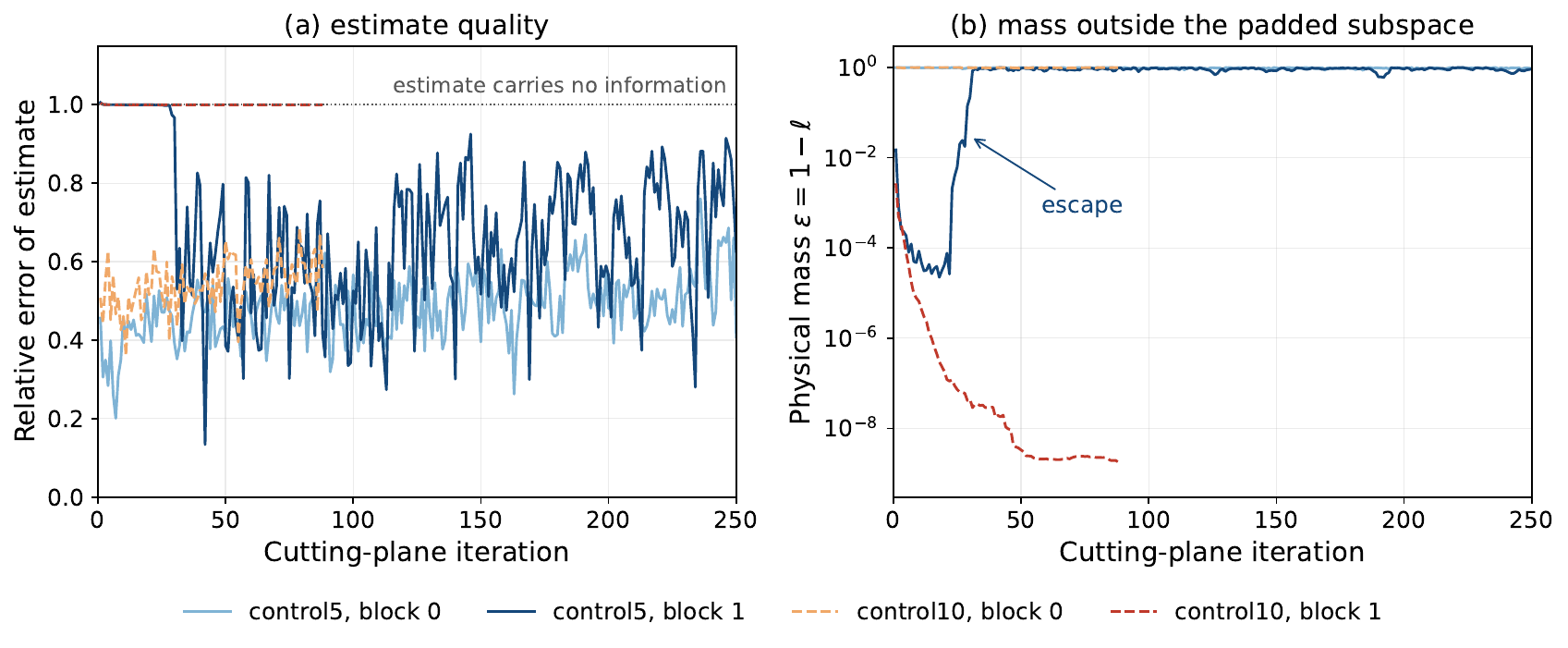}
\caption{Quality of the VQE eigenvalue estimates. (a) Relative error against cutting-plane iteration. (b) Physical mass $\epsilon = 1-\ell$, on a logarithmic scale. On \texttt{control10} the smaller block descends to a floor near $10^{-9}$ and does not recover; on \texttt{control5} it floors four orders of magnitude higher and escapes at iteration $23$. Curves are from the first repeat.}
\label{fig:estimates}
\end{figure}

Two regimes are visible (Figure~\ref{fig:estimates}). In the first, the estimate carries the correct sign and a substantial fraction of the magnitude: relative errors settle at $0.43$ to $0.54$ on the larger block of \texttt{control3}--\texttt{control10} --- and higher on the two smallest instances ---, with median leaked amplitudes below $0.06$. In the second, the optimized state collapses almost entirely into the padded subspace, and thus, the energy estimate approaches zero from above. This occurs on the smaller~block of \texttt{control7}--\texttt{control10}, where the leaked amplitude exceeds $0.997$ at every iteration, no cut is ever produced, and the oracle falsely certifies the block PSDness in all separation calls, against median exact minimum eigenvalues of $-35$ to $-38$ and individual values ranging from $-18.6$ to $-53.0$. 
In these cases, the cutting plane scheme does not terminate spuriously because the larger block continues to supply violated cuts. 

The near-zero estimates on these blocks are not small eigenvalues reported accurately. The estimate factors as $\epsilon R$, where $R$ is the Rayleigh quotient of $X^*$ at the surviving direction --- the value the estimate would take if the state carried no leakage. Since $R$ is invariant under rescaling of the truncated vector $u$, it distinguishes whether an estimate is near zero due to a genuinely small eigenvalue or due to the state leaking almost entirely into the padded subspace. On \texttt{control7}--\texttt{control10} it is the latter: $\epsilon$ falls to $10^{-6}$ or below while $R$ has per-instance medians of $+24$ to $+68$, against minimum eigenvalues near $-35$ to $-38$. The surviving component carries a positive Rayleigh quotient, far above the minimum, so no rescaling of it results in a better estimate. Note that the larger block of the same instance is padded to the same fraction of its dimension yet has negative $R$ throughout and never collapses, so the collapse is not a consequence of the amount of padding.

\paragraph{Quality of the bound} Expressed as a fraction of the initial optimality gap, the exact oracle closes $75$ to $82\%$ on every instance in the family, with a standard deviation of $2.7$ percentage points and no particular trend in instance size. The variational oracle trails it by less than $8$ percentage points on \texttt{control1}--\texttt{control5}, closing $72$ to $82\%$, and then decreases monotonically. The performance of the cutting-plane scheme under an exact oracle is thus essentially size-independent over this family.

The degradation steepens at \texttt{control6} and continues from there. Part of it coincides with the reduction in the iteration budget, across which the exact oracle also loses $5$ percentage points of closed gap; the VQE oracle plausibly loses more, its cuts being individually weaker. From \texttt{control7} onward, the collapse compounds this, and the smaller block is never separated: all cuts originate from the larger block, while the exact oracle separates both. The effect is visible in the cut counts of Table~\ref{tab:scaling}. Second-order-cone cuts, requiring two violating directions, drop to exactly one per iteration on \texttt{control7}--\texttt{control10}, half what the exact oracle generates, since only one block supplies a direction. Linear cuts are also fewer, by $12$ to $36\%$ on \texttt{control2}--\texttt{control10}, the directions collected along a single optimizer trajectory being correlated and largely removed by the diversity filter of Section~\ref{sec:cut_generation}. The exception is \texttt{control1}, where the two oracles generate the same number to within $1\%$. On \texttt{control9} and \texttt{control10} the two runs reached different iteration counts, and the per-iteration reductions are $31\%$ and $38\%$ respectively, larger than the totals suggested by Table~\ref{tab:scaling}.

On \texttt{control1}, whose blocks require only four and three qubits, the variational oracle attains a bound at least as good as the exact one ($66.2 \pm 6.6$ against $71.8$, over three repeats). A possible explanation is that the ansatz is expressive enough at this size to return useful directions beyond the eigenvectors of negative eigenvalues, to which the exact oracle is restricted. The effect does not survive to larger blocks.

\paragraph{Padding the null subspace} Zero-padding places the padded eigenvalues at $0$, above $\lambda_{\min}(X^*)$; padding with $\delta I$ for $\delta \ge \lambda_{\max}(X^*)$ places them above the entire spectrum instead, so that mass in the padded subspace can no longer be traded against the physical component to lower the objective. Cuts remain valid, since $\tilde u^\top X' \tilde u = u^\top X^* u + \delta\|w\|^2$ with $\delta\|w\|^2 \ge 0$, so $u^\top X^* u < 0$ whenever $\tilde u^\top X' \tilde u$ is negative; \eqref{eq:truncation} thus holds as an inequality rather than an equality. Taking $\delta$ to be the Gershgorin bound \cite{gershgorin1931uber} keeps the cost at $O(n^2)$. This remedy removes the collapse but does not improve the bounds. On \texttt{control7} at $50$ iterations under matched seeds, the leaked amplitude on the smaller block falls from nearly one to $0.0004$ and the block supplies cuts in $43$ of $50$ iterations, against none under zero-padding. The master objective is nonetheless worse, $2527$ against $2108$. Two effects account for this. First, with leakage suppressed, the collected states no longer differ significantly in their padded coordinates, so the cosine-similarity test of Section~\ref{sec:cut_generation} rejects $86\%$ of candidates on the larger block against $79\%$ before, filtering out some of the physical directions that would otherwise be admitted. Second, applying the correction to the larger block is ultimately counterproductive: the evaluation budget goes to suppressing a leakage below $1\%$ rather than to resolving the eigenvalue, and the relative error of the estimate rises from $0.46$ to $0.65$. We therefore retain zero-padding in the results reported above, noting that applying the correction only to the blocks that collapse would be the natural refinement.

\paragraph{Memory} The scheme was motivated in part by the potentially low memory requirements of solving SDPs at scale, so we also examine how memory is used. We sample the resident set size (RSS) of the solver process --- the physical memory currently held --- four times per cutting-plane iteration: before and after each separation call, and before and after the master solve.

The median memory retained across a separation call is virtually zero under both oracles: even on \texttt{control10}, the largest instance and the one at which memory was exhausted, the samples taken before and after separation are within a megabyte. Across a master solve, by contrast, the median increase rises from under a megabyte per iteration on the smaller instances to $87$ megabytes on \texttt{control10}, and peak usage follows accordingly, from $157$ megabytes on \texttt{control1} to the $8.4$ gigabytes at which the run was terminated. This asymmetry is structural rather than an artifact of simulating small registers: a $q$-qubit state vector holds $2^q = N$ amplitudes, fewer than the $O(N^2)$ numbers required to store the block itself, whereas the master problem accumulates cuts together with the working memory the conic solver needs to factor and solve the resulting relaxations. Both are classical, and both are present regardless of how the separation subproblem is solved.

Two conclusions follow. First, the memory bottleneck of the method is not the separation step: the two oracles reach comparable peak use (Figure~\ref{fig:scaling}b), differing by $4\%$ to $41\%$ in both directions on \texttt{control2}--\texttt{control10} with no systematic advantage to either; the largest discrepancy is on \texttt{control1}, where both figures are small in absolute terms. Second, the exact oracle exhausted memory on two instances against one for the variational oracle, and reached fewer iterations on both, consistent with the cut counts of Table~\ref{tab:scaling}: the variational oracle extends its reach by generating cuts more sparingly, not by separating more cheaply. We return to the implications for the space-complexity motivation in Section~\ref{sec:discussion}.

\subsection{Encoding cost} \label{sec:encoding-cost}
\begin{table}[htbp]\footnotesize
\caption{Pauli decomposition of the padded candidate matrices across the family, first repeat. For each block, $n$ is its dimension, $q = \lceil \log_2 n \rceil$ the register size, ``Pauli terms'' the range of Pauli term counts observed across cutting-plane iterations, and ``Med.'' the median count as a fraction of the $N(N+1)/2$ strings available to a real symmetric operator on $N = 2^q$ coordinates. The induced entanglement map was the complete graph on all $q$ qubits at every iteration of every run.}
\label{tab:pauli-density}
\centering
\begin{tabular}{l rcrr rcrr}
\hline
 & \multicolumn{4}{c}{Block 0} & \multicolumn{4}{c}{Block 1} \\
\cline{2-5}\cline{6-9}
Instance & $n$ & $q$ & Pauli terms & Med. & $n$ & $q$ & Pauli terms & Med. \\
\hline
\texttt{control1}  &  10 & 4 &  120--136  & $100.0\%$ &  5 & 3 &   36       & $100.0\%$ \\
\texttt{control2}  &  20 & 5 &  464--528  & $100.0\%$ & 10 & 4 &  136       & $100.0\%$ \\
\texttt{control3}  &  30 & 5 &  524--528  & $100.0\%$ & 15 & 4 &  136       & $100.0\%$ \\
\texttt{control4}  &  40 & 6 & 1802--2080 &  $99.5\%$ & 20 & 5 &  526--528  & $100.0\%$ \\
\texttt{control5}  &  50 & 6 & 2055--2078 &  $99.5\%$ & 25 & 5 &  528       & $100.0\%$ \\
\texttt{control6}  &  60 & 6 & 2066--2079 &  $99.7\%$ & 30 & 5 &  528       & $100.0\%$ \\
\texttt{control7}  &  70 & 7 & 6595--8162 &  $98.0\%$ & 35 & 6 & 2072--2080 & $100.0\%$ \\
\texttt{control8}  &  80 & 7 & 7122--8203 &  $98.8\%$ & 40 & 6 & 2080       & $100.0\%$ \\
\texttt{control9}  &  90 & 7 & 8146--8205 &  $99.0\%$ & 45 & 6 & 2079--2080 & $100.0\%$ \\
\texttt{control10} & 100 & 7 & 8159--8197 &  $99.1\%$ & 50 & 6 & 2079--2080 & $100.0\%$ \\
\hline
\end{tabular}
\end{table}
Table~\ref{tab:pauli-density} reports the Pauli decomposition of the padded candidate matrices across the scaling study. The decomposition is essentially dense: the median term count is at least $98\%$ of the $N(N+1)/2$ strings available to a real symmetric operator on every block of every instance, and exactly $100\%$ on $14$ of the $20$ blocks. The sparsest decompositions occur at the first iteration on the larger block, falling to $79.9\%$ on \texttt{control7}, but the count exceeds $95\%$ from the second iteration onward in every run. This is not surprising given how the candidate matrices arise: the master problem's optimal solutions are dense, and each cut added is a dense
rank-one constraint.

Two consequences follow. The cost of a cost function evaluation is fixed by the register size alone: with the decomposition dense, the number of terms to be estimated is $N(N+1)/2 \approx 4^{q}/2$ regardless of the instance, which is why separation time in Figure~\ref{fig:scaling}a is constant within each group of blocks sharing a value of $q$ and steps by a factor of $2.4$ to $3.5$ per additional qubit. Moreover, as observed in the configuration experiments of Section~\ref{sec:vqe-config}, the induced entanglement map is the complete graph on all $q$ qubits at every iteration examined, so ODA effectively yields full entanglement across the entire family. We return to the implications in Section~\ref{sec:discussion}.

\subsection{Finite sampling and device noise}\label{sec:noisy}

All results so far were obtained with expectation values computed exactly by state vector simulation. We now examine what happens when they are estimated from a finite number of measurements, and when the circuits are additionally executed under a device noise model. To this end, we carried out two experiments. The first isolates the separation oracle: candidate matrices are saved at selected iterations of a noiseless \texttt{control2} run and the oracle is called on each of them directly, outside the cutting-plane loop, across a grid of execution modes and shot counts. The second embeds the noisy oracle in the full scheme and compares the resulting run against the noiseless one.

\subsubsection{Oracle accuracy}

Candidate matrices were saved from a $250$-iteration noiseless run, at iterations $1$, $25$, $50$, $100$ and $250$ --- both blocks at each, so ten matrices --- together with the exact minimum eigenvalue of each. The oracle was then invoked on each saved matrix under three execution modes: (1) full state vector simulation, (2) ideal sampling on \texttt{aer\_simulator}, and (3) sampling with the \texttt{ibm\_fake\_fez} device noise model, both (2) and (3) at $1024$, $4096$ and $8192$ shots. Three repeats were performed per grid point, resulting in a total of $210$ separation calls. Each call was cold-started and made with multi-cut collection disabled, so that what is measured is the estimate itself rather than the loop's ability to recover from it. Results are given in Table~\ref{tab:noisy-oracle}.

\begin{table}[htbp]\footnotesize
\caption{Separation oracle under finite sampling and device noise, on candidate matrices saved from a noiseless \texttt{control2} run. ``Neg.\ est.'' is the fraction of the $30$ calls in each row returning a negative estimate; every saved block has a negative minimum eigenvalue, so a positive estimate is a false PSDness certificate at the block level. Relative error, leaked amplitude and evaluation count are medians over the $30$ calls; the last two columns give the relative error restricted to the matrices saved at iteration $1$ and $250$ averaged over the two blocks.}
\begin{tabular}{lr|rrrr|rr}
\hline
 & & & & & & \multicolumn{2}{c}{Rel.\ err.} \\
Mode & Shots & Neg.\ est. & Rel.\ err. & Leaked\ ampl. & Eval.\ count & iter.\ 1 & iter.\ 250 \\
\hline
noiseless & ---  & $83\%$ & 0.562 & 0.179 & 200 & 0.25 & 0.90 \\
sampling  & 1024 & $37\%$ & 1.111 & 0.543 & 150 & 0.89 & 3.59 \\
sampling  & 4096 & $40\%$ & 1.038 & 0.616 & 148 & 0.98 & 1.29 \\
sampling  & 8192 & $47\%$ & 1.005 & 0.464 & 154 & 0.86 & 1.42 \\
noisy     & 1024 & $37\%$ & 1.425 & 0.439 & 148 & 0.82 & 2.77 \\
noisy     & 4096 & $20\%$ & 1.415 & 0.593 & 152 & 0.80 & 3.93 \\
noisy     & 8192 & $23\%$ & 1.417 & 0.491 & 156 & 0.71 & 4.94 \\
\hline
\end{tabular}
\label{tab:noisy-oracle}
\end{table}

The estimate degrades in two stages. Finite sampling alone reduces the fraction of calls returning a negative estimate from $83\%$ to $41\%$ and raises the median relative error above unity; adding the device noise model reduces it further to $27\%$. The leaked amplitude rises from a median of $0.18$ to roughly $0.5$ in both sampled modes, and the optimizer stops before exhausting its evaluation budget, at a median of about $150$ evaluations against $200$ noiseless. Separation time rises by a factor of three under sampling and fourteen under the noise model, at fixed evaluation budget.

Increasing the shot count does not recover the loss. Comparing $1024$ against $8192$ shots within each combination of saved matrix and block, so that the comparison is not confounded by the differing eigenvalue magnitudes across snapshots, the absolute error improves in eight of $10$ cases under ideal sampling, with a median ratio of $1.25$. This is an improvement, but far short of the factor of $\sqrt{8} \approx 2.83$ that would follow if sampling variance were the dominant error. The noiseless median relative error of $0.562$ is a floor set by the choice of ansatz and the optimizer, and eightfold more shots cannot go below it. Under the device noise model, the same comparison improves in six of $10$ cases with a median ratio of $1.02$, i.e., no improvement resolvable at this sample size, as expected of an error that does not shrink with the number of~measurements.

The failures are concentrated where oracle precision becomes more critical. As the cutting-plane iterate approaches the PSD cone, the minimum eigenvalues of the saved matrices shrink, while the oracle's relative error does not decline proportionately. Averaged over both blocks, the relative error at iteration $1$ ranges from $0.26$ under exact simulation to $0.98$ under sampling, and rises further to between $0.90$ and $4.94$ by iteration $250$ across every mode (Table~\ref{tab:noisy-oracle}). This is the behavior anticipated in Section~\ref{sec:cost_func_eval}: the required accuracy of the
oracle grows as the scheme refines the outer approximation, and every source of inexactness we examine --- finite sampling, device noise, ansatz, optimizer --- degrades in the same direction.

\subsubsection{Behavior within the cutting-plane loop}

We also ran the full scheme on \texttt{control2} with the oracle executing under the \texttt{ibm\_fake\_fez} noise model at $4096$ shots, for $100$ iterations and three repeats, against the noiseless run truncated at the same iteration. Results are summarized in Table~\ref{tab:noisy-loop}.

\begin{table}[htbp]\footnotesize
\caption{The cutting-plane scheme on \texttt{control2} with a noisy separation oracle (\texttt{ibm\_fake\_fez}, $4096$ shots), against the noiseless run at the same iteration count. ``Candidates'' is the number of evaluations across the run whose estimator reading was negative, and ``Rejected'' the percentage of those whose quadratic form $u^\top X^* u$, computed classically from the state amplitudes, turned out to be non-negative. Repeat $2$ terminated early with a false PSDness certificate. }
\label{tab:noisy-loop}
\centering
\begin{tabular}{lrrrrrr}
\hline
Run & Iterations & Bound & Linear cuts & Sep./iter.\ (s) & Candidates & Rejected \\
\hline
noiseless        & 100 & 210.6 & 1262 &  7.0 & --- & --- \\
noisy, repeat 1  & 100 & 377.5 &  501 & 74.7 & 8518 & $16.0\%$ \\
noisy, repeat 2  &  90 & 396.0 &  490 & 74.7 & 9107 & $11.5\%$ \\
noisy, repeat 3  & 100 & 378.4 &  573 & 74.7 & 9864 & $12.2\%$ \\
\hline
\end{tabular}
\end{table}

Three observations follow. First, one of the three repeats terminated with a false PSDness certificate at iteration $90$ as the oracle reported both blocks PSD while an exact eigendecomposition of the same iterate identified negative minimum eigenvalues. 

Second, the loop separates reliably on one block and not on the other. The larger block was separated in every one of the $290$ calls across the three repeats, with an average of $70$ evaluations per call returning a negative estimator reading, of which $57$ also had a negative quadratic form when checked classically. On the smaller block no candidate was collected in $79\%$ of calls, against a true minimum eigenvalue with median $-6.49$; in the remaining $21\%$ it was separated. As in Section~\ref{sec:main-results}, the block that fails is the one whose minimum eigenvalue is further from zero.

Third, noisy separation necessitates a classical check on collected candidates. The estimator and the exact quadratic form $u^\top X^* u$ agree to machine precision under state vector simulation, where the check rejects a negligible $0.002\%$ of the candidates. Nevertheless, it rejects 11.5-16.0\% under the noise model across the three repeats (Table~\ref{tab:noisy-loop}). These are states whose energy estimate is negative but whose exact quadratic form is not, and consequently, the cuts they would produce are not violated~at the current iterate. Note here that this check requires the amplitudes of the collected states, which are available in simulation but would require state tomography on hardware.

Overall, noise incurs a substantial cost to the cutting plane scheme: (1) the bound at iteration $100$ is $377.5$ to $396.0$ against $210.6$ in the noiseless scenario, (2) the number of linear cuts separated decreases by more than half, and (3) separation times are an order of magnitude higher under the noise model.

\subsection{Hardware evaluation} \label{sec:hardware}

The experiments of Section~\ref{sec:noisy} model device error with a noise model. Next, we execute the same separation oracle on a physical device, to check whether the behavior observed under noisy simulations reproduces on hardware. For this experiment, candidate matrices were saved from a noiseless \texttt{control1} run at iterations $1$, $25$, $100$ and $250$, both blocks at each, and the oracle was called on each of the resulting eight matrices under three execution modes: exact state vector simulation, the \texttt{ibm\_fake\_fez} noise model, and \texttt{ibm\_kingston} hardware, the latter two at $1024$ shots. The simulated modes were repeated three times per matrix, the hardware runs once, yielding $24$, $24$ and $8$ separation calls respectively. As in Section~\ref{sec:noisy}, each call is cold-started with multi-cut collection disabled. \texttt{control1} is the smallest instance in the family, its two blocks requiring four and three qubits; the circuits submitted have transpiled depths of $100$ and $67$ with $36$ and $21$ two-qubit gates. The total quantum processor time consumed was $7.4$ seconds. Results are reported in Table~\ref{tab:hardware}.

\begin{table}[htbp]\footnotesize
\caption{Separation oracle on \texttt{control1} candidate matrices, under exact simulation, a device noise model, and hardware. Each cell reports the number of calls returning a negative estimate out of those performed, with the median relative error in parentheses; both blocks are pooled at each iteration. Every saved block has a negative minimum eigenvalue, so a positive estimate is a false PSDness certificate at the block level.}
\label{tab:hardware}
\centering
\begin{tabular}{lcccc}
\hline
 & \multicolumn{4}{c}{Saved at cutting-plane iteration} \\
\cline{2-5}
Mode & $1$ & $25$ & $100$ & $250$ \\
\hline
exact simulation             & $6/6$ \ (0.03) & $4/6$ \ (0.61) & $5/6$ \ (0.77) & $6/6$ \ (0.52) \\
\texttt{ibm\_fake\_fez}, 1024 shots & $4/6$ \ (0.87) & $2/6$ \ (7.01) & $1/6$ \ (2.99) & $0/6$ \ (14.84) \\
\texttt{ibm\_kingston}, 1024 shots & $2/2$ \ (0.31) & $0/2$ \ (4.28) & $0/2$ \ (2.78) & $0/2$ \ (3.48) \\
\hline
\end{tabular}
\end{table}

The results indicate that at these block dimensions, exact simulation does not degrade along the trajectory: it returns a negative estimate in $21$ of $24$ calls, and in all six calls on the matrices saved at the last iteration. This differs from \texttt{control2} in Section~\ref{sec:noisy}, where the ansatz and the optimizer were by themselves enough to lose the negative sign at the later iterates. Here they are not, so the degradation in the other two rows is attributable to finite sampling and device error rather than to the variational method itself.

In terms of the number of negative eigenvalue estimates, hardware and the noise model exhibit closely aligned behavior. Sign correctness is $25\%$ on hardware against $29\%$ under \texttt{ibm\_fake\_fez}. Both return a negative estimate for the blocks saved at the first iteration and fail on most or all of the later ones. Within the resolution afforded by a single hardware repeat per matrix, the noise model serves as a reasonable proxy for the device.

We do not perform a hardware execution of the full scheme because forming a cut requires the amplitudes of the optimized state. On hardware, recovering these requires state tomography, which has a cost exponential in the number of qubits. The classical verification of collected candidates described in Section~\ref{sec:noisy} is unavailable for the same reason. An end-to-end hardware realization of the cutting plane scheme therefore awaits a cheaper readout scheme for cut generation.

\section{Discussion \& concluding remarks}\label{sec:discussion}
We have proposed a cutting-plane framework for semidefinite programming in which the
separation oracle is realized by a variational quantum eigensolver, and have studied it
computationally on the \texttt{control} family of SDPLIB under exact simulation, finite
sampling, a device noise model, and on quantum hardware. The framework runs end-to-end and
the variational oracle produces valid cuts throughout, closing a substantial fraction of
the initial optimality gap on every instance and matching the exact oracle on the smallest.
To our knowledge this is the first study to place a variational eigensolver inside a
cutting-plane loop for SDP and to report its behavior across this range of execution
conditions.

The property that motivates the construction is the qubit requirement: estimating the
extremal eigenpair of an $N \times N$ operator needs a register of $q = \lceil \log_2
N\rceil$ qubits, against the $O(N^2)$ numbers required to store the operator classically.
Our results indicate that this logarithmic scaling, while real, is not by itself sufficient
to yield an advantage within a cutting-plane scheme, for four reasons that the study
isolates.

\emph{The memory bottleneck is classical.} The scheme was motivated in part by the memory required by checking PSDness
at scale, but the peak memory consumption in our implementation is not due to the separation step. The median memory retained across a separation call is within a megabyte on every instance under either oracle, while the master problem retains up to $87$ megabytes per iteration on \texttt{control10}, and both oracles reach the same
memory wall. The accumulating cut pool and the working
memory of the conic solver are classical costs, present regardless of how the separation subproblem is solved.

\emph{The encoding cost is exponential in the register size.} The qubit count is logarithmic, but the number of Pauli terms whose expectation must be estimated is not: the
candidate matrices are essentially dense in the Pauli basis, with median term counts at or above $98\%$ of the $N(N+1)/2$ strings available on every block of every instance. The per-evaluation cost is therefore $\Theta(4^q)$ in practice, and the measured separation
times form four plateaus indexed by $q$, rising by a factor of $2.4$ to $3.5$ per additional qubit. The same measurement removes the premise of the operator-derived ansatz we introduced: the entanglement map it induces is the complete graph at every iteration examined, so the construction reduces to full entanglement across the family.

\emph{Cut generation requires more than an eigenvalue estimate.} Forming a cut requires the amplitudes of the optimized state, not merely its expectation value. In simulation these are free; on hardware they would require state tomography, whose cost is exponential in the number of qubits. This is why our hardware evaluation characterizes the oracle at candidate matrices sampled from a cutting-plane trajectory rather than executing the scheme end to end, and it is the principal obstacle to a hardware realization. The classical verification of collected candidates that proves necessary under noise is unavailable for the same reason.

\emph{The outer scheme binds before the oracle's precision does.} Even with an exact oracle, the polyhedral scheme converges slowly: after $500$ iterations on \texttt{control2} the best configuration attains a bound of $88.24$ against an optimal value of $8.30$, and the exact oracle closes a near-constant $75$--$82\%$ of the initial gap on every instance, with no trend in size. Improvements to the separation oracle therefore act on what is not the limiting factor. 

Two findings about the oracle itself are worth restating. The first is a failure mode introduced by the embedding rather than by the quantum method: padding a block to the next power of two places a null eigenspace at zero, above $\lambda_{\min}(X^*)$ and therefore locally attractive, and on the smaller block of the four largest instances the optimized state collapses into it at every iteration, falsely certifying a block as PSD while its minimum eigenvalue is near $-38$. Padding with $\delta I$ for $\delta \ge \lambda_{\max}(X^*)$ removes the collapse at no cost in cut validity or encoding size, but does not improve the bound, the recovered cuts being fewer and individually weaker. The second is that inexactness degrades as the iterate approaches the PSD cone. Under a device noise model this produces false PSDness certificates; hardware reproduces this behavior closely enough at the scale we performed our tests.

Several directions for future research follow. The padding correction is worth applying selectively, to the blocks that collapse rather than uniformly, since elsewhere it consumes optimizer budget to no purpose. In addition, a readout scheme yielding a violating direction without full tomography would remove the obstacle to hardware execution and is the prerequisite for any practical realization. Beyond the oracle, the outer approximation is what limits convergence, so stronger valid inequalities would benefit both oracles alike. Finally, since what blocks hardware execution is the readout rather than the eigenvalue estimate, a more promising point of insertion may be the step-length computation of an interior-point method, which is itself a minimum eigenvalue problem but one that uses only the eigenvalue and discards the eigenvector, and so requires no tomography.

\subsection*{Acknowledgments}
We thank Cihan Okay for helpful discussions, in particular regarding the VQE-based multi-cut generation strategy. The authors acknowledge the use of Claude AI as an interactive assistant to help draft and refine the manuscript text. All AI-generated content were carefully reviewed, revised, and validated by the authors. Additionally, the tool aided in implementing and debugging parts of the code, processing and analyzing numerical results, and writing scripts for figure generation. The authors take full responsibility for the integrity and accuracy of the final manuscript and its results.

\newpage

\bibliographystyle{siamplain}
\bibliography{references}

\begin{thebibliography}{10}

\bibitem{apers2023quantum}
{\sc S.~Apers and S.~Gribling}, {\em Quantum speedups for linear programming via interior point methods}, arXiv preprint arXiv:2311.03215,  (2023).

\bibitem{augustino2023quantum}
{\sc B.~Augustino, J.~Leng, G.~Nannicini, T.~Terlaky, and X.~Wu}, {\em A quantum central path algorithm for linear optimization}, arXiv preprint arXiv:2311.03977,  (2023).

\bibitem{augustino2023quantumSDP}
{\sc B.~Augustino, G.~Nannicini, T.~Terlaky, and L.~F. Zuluaga}, {\em Quantum interior point methods for semidefinite optimization}, Quantum, 7 (2023), p.~1110.

\bibitem{augustino2021inexactsocp}
{\sc B.~Augustino, T.~Terlaky, M.~Mohammadisiahroudi, and L.~F. Zuluaga}, {\em An inexact-feasible quantum interior point method for second-order cone optimization}, quantum, 16 (2021), p.~29.

\bibitem{bai2008semidefinite}
{\sc X.~Bai, H.~Wei, K.~Fujisawa, and Y.~Wang}, {\em Semidefinite programming for optimal power flow problems}, Int. J. Electr. Power Energy Syst., 30 (2008), pp.~383--392.

\bibitem{bertvempala2004_convexprograms_randomwalks}
{\sc D.~Bertsimas and S.~Vempala}, {\em Solving convex programs by random walks}, J. ACM, 51 (2004), p.~540–556, \url{https://doi.org/10.1145/1008731.1008733}.

\bibitem{borchers1999sdplib}
{\sc B.~Borchers}, {\em Sdplib 1.2, a library of semidefinite programming test problems}, Optim. Methods Softw., 11 (1999), pp.~683--690.

\bibitem{brandao2017quantum}
{\sc F.~G. Brand{\~a}o, A.~Kalev, T.~Li, C.~Y.-Y. Lin, K.~M. Svore, and X.~Wu}, {\em Quantum sdp solvers: Large speed-ups, optimality, and applications to quantum learning}, arXiv preprint arXiv:1710.02581,  (2017).

\bibitem{brandao2017quantumspeedup}
{\sc F.~G. Brandao and K.~M. Svore}, {\em Quantum speed-ups for solving semidefinite programs}, in 2017 IEEE 58th Annual Symposium on Foundations of Computer Science (FOCS), IEEE, 2017, pp.~415--426.

\bibitem{bernal2024copositivehybrid}
{\sc R.~Brown, D.~E. Bernal~Neira, D.~Venturelli, and M.~Pavone}, {\em A copositive framework for analysis of hybrid ising-classical algorithms}, SIAM J. Optim., 34 (2024), pp.~1455--1489, \url{https://doi.org/10.1137/22M1514581}.

\bibitem{cerezo2021variational}
{\sc M.~Cerezo, A.~Arrasmith, R.~Babbush, S.~C. Benjamin, S.~Endo, K.~Fujii, J.~R. McClean, K.~Mitarai, X.~Yuan, L.~Cincio, et~al.}, {\em Variational quantum algorithms}, Nat. Rev. Phys., 3 (2021), pp.~625--644.

\bibitem{chen2025slackSDP}
{\sc J.~Chen, H.~Westerheim, Z.~Holmes, I.~Luo, T.~Nuradha, D.~Patel, S.~Rethinasamy, K.~Wang, and M.~M. Wilde}, {\em Slack-variable approach for variational quantum semidefinite programming}, Phys. Rev. A, 112 (2025), p.~022607, \url{https://doi.org/10.1103/lwxq-4myj}.

\bibitem{ciacco2026cuttingplanemethodologyquantumoptimization}
{\sc A.~Ciacco, L.~D.~P. Pugliese, and F.~Guerriero}, {\em Cutting-plane methodology via quantum optimization for solving the traveling salesman problem}, 2026, \url{https://arxiv.org/abs/2604.20321}.

\bibitem{de2009semidefinite}
{\sc E.~De~Klerk, D.~V. Pasechnik, and R.~Sotirov}, {\em On semidefinite programming relaxations of the traveling salesman problem}, SIAM J. Optim., 19 (2009), pp.~1559--1573.

\bibitem{ellinas2024hybridquantumclassicalalgorithmmixedinteger}
{\sc P.~Ellinas, S.~Chevalier, and S.~Chatzivasileiadis}, {\em A hybrid quantum-classical algorithm for mixed-integer optimization in power systems}, 2024, \url{https://arxiv.org/abs/2404.10693}.

\bibitem{gershgorin1931uber}
{\sc S.~A. Gershgorin}, {\em Uber die abgrenzung der eigenwerte einer matrix}, Известия Российской академии наук. Серия математическая,  (1931), pp.~749--754.

\bibitem{Goemans1995}
{\sc M.~Goemans and D.~Williamson}, {\em Improved approximation algorithms for maximum cut and satisfiability problems using semidefinite programming}, Journal of the ACM, 42 (1995), pp.~1115--1145, \url{https://doi.org/10.1145/227683.227684}.

\bibitem{kandala2017hardware}
{\sc A.~Kandala, A.~Mezzacapo, K.~Temme, M.~Takita, M.~Brink, J.~M. Chow, and J.~M. Gambetta}, {\em Hardware-efficient variational quantum eigensolver for small molecules and quantum magnets}, nature, 549 (2017), pp.~242--246.

\bibitem{kerenidis2020quantum}
{\sc I.~Kerenidis and A.~Prakash}, {\em A quantum interior point method for lps and sdps}, ACM Trans. Quantum Comput., 1 (2020), pp.~1--32.

\bibitem{krishnan2006unifying}
{\sc K.~Krishnan and J.~E. Mitchell}, {\em A unifying framework for several cutting plane methods for semidefinite programming}, Optim. Methods Softw., 21 (2006), pp.~57--74.

\bibitem{lasserre2001global}
{\sc J.~B. Lasserre}, {\em Global optimization with polynomials and the problem of moments}, SIAM J. Optim., 11 (2001), pp.~796--817.

\bibitem{Le2024VQEconstrained}
{\sc T.~V. Le and V.~Kekatos}, {\em Solving constrained optimization problems via the variational quantum eigensolver with constraints}, Phys. Rev. A, 110 (2024), p.~022430, \url{https://doi.org/10.1103/PhysRevA.110.022430}.

\bibitem{le2026solvingconicprogramssparse}
{\sc T.~V. Le, M.~M. Wilde, and V.~Kekatos}, {\em Solving conic programs over sparse graphs using a variational quantum approach: The case of the optimal power flow}, 2026, \url{https://arxiv.org/abs/2509.00341}.

\bibitem{liu2022integrated}
{\sc F.~Liu, Y.~Cui, C.~Masouros, J.~Xu, T.~X. Han, Y.~C. Eldar, and S.~Buzzi}, {\em Integrated sensing and communications: Toward dual-functional wireless networks for 6g and beyond}, IEEE journal on selected areas in communications, 40 (2022), pp.~1728--1767.

\bibitem{liu2025quantumthermodynamicssemidefiniteoptimization}
{\sc N.~Liu, M.~Minervini, D.~Patel, and M.~M. Wilde}, {\em Quantum thermodynamics and semi-definite optimization}, 2025, \url{https://arxiv.org/abs/2505.04514}.

\bibitem{lovasz2003semidefinite}
{\sc L.~Lov{\'a}sz}, {\em Semidefinite programs and combinatorial optimization}, in Recent advances in algorithms and combinatorics, Springer, 2003, pp.~137--194.

\bibitem{luo2010semidefinite}
{\sc Z.-Q. Luo, W.-K. Ma, A.~M.-C. So, Y.~Ye, and S.~Zhang}, {\em Semidefinite relaxation of quadratic optimization problems}, IEEE Signal Processing Magazine, 27 (2010), pp.~20--34.

\bibitem{marecek2021cutting}
{\sc J.~Marecek and A.~Akhriev}, {\em A cutting-plane method for semidefinite programming with potential applications on noisy quantum devices}, arXiv preprint arXiv:2110.03400,  (2021).

\bibitem{mcclean2018barren}
{\sc J.~R. McClean, S.~Boixo, V.~N. Smelyanskiy, R.~Babbush, and H.~Neven}, {\em Barren plateaus in quantum neural network training landscapes}, Nature communications, 9 (2018), p.~4812.

\bibitem{mohammadisiahroudi2025quantum}
{\sc M.~Mohammadisiahroudi, B.~Augustino, P.~Sampourmahani, and T.~Terlaky}, {\em Quantum computing inspired iterative refinement for semidefinite optimization}, Math. Program.,  (2025), pp.~1--40.

\bibitem{mohammadisiahroudi2022efficient}
{\sc M.~Mohammadisiahroudi, R.~Fakhimi, and T.~Terlaky}, {\em Efficient use of quantum linear system algorithms in interior point methods for linear optimization}, arXiv preprint arXiv:2205.01220,  (2022).

\bibitem{nannicini2024fastLP}
{\sc G.~Nannicini}, {\em Fast quantum subroutines for the simplex method}, Operations Research, 72 (2024), pp.~763--780.

\bibitem{nesterov1994interior}
{\sc Y.~Nesterov and A.~Nemirovskii}, {\em Interior-point polynomial algorithms in convex programming}, SIAM, 1994.

\bibitem{Nie2026quantumalternating}
{\sc H.~Nie, D.~An, and Z.~Wen}, {\em Quantum {A}lternating {D}irection {M}ethod of {M}ultipliers for {S}emidefinite {P}rogramming}, {Quantum}, 10 (2026), p.~2154, \url{https://doi.org/10.22331/q-2026-07-08-2154}.

\bibitem{parrilo2003semidefinite}
{\sc P.~A. Parrilo}, {\em Semidefinite programming relaxations for semialgebraic problems}, Math. Program., 96 (2003), pp.~293--320.

\bibitem{patel2024variational}
{\sc D.~Patel, P.~J. Coles, and M.~M. Wilde}, {\em Variational quantum algorithms for semidefinite programming}, Quantum, 8 (2024), p.~1374.

\bibitem{pena2007computing}
{\sc J.~Pena, J.~Vera, and L.~F. Zuluaga}, {\em Computing the stability number of a graph via linear and semidefinite programming}, SIAM J. Optim., 18 (2007), pp.~87--105.

\bibitem{peng2025hybridquantumbranchandboundmethod}
{\sc Z.~Peng, D.~de~Roux, and D.~E.~B. Neira}, {\em Hybrid quantum branch-and-bound method for quadratic unconstrained binary optimization}, 2025, \url{https://arxiv.org/abs/2509.11040}.

\bibitem{skrzypczyk2023semidefinite}
{\sc P.~Skrzypczyk and D.~Cavalcanti}, {\em Semidefinite programming in quantum information science}, IOP Publishing, 2023.

\bibitem{sun2026hybridquantumclassicalbranchandpriceintraday}
{\sc P.~Sun, L.~Zhong, Q.-G. Zeng, and L.~Wang}, {\em Hybrid quantum-classical branch-and-price for intra-day electric vehicle charging scheduling via partition coloring}, 2026, \url{https://arxiv.org/abs/2603.21374}.

\bibitem{todd2001semidefinite}
{\sc M.~J. Todd}, {\em Semidefinite optimization}, Acta Numerica, 10 (2001), pp.~515--560.

\bibitem{tunccel2016polyhedral}
{\sc L.~Tun{\c{c}}el}, {\em Polyhedral and semidefinite programming methods in combinatorial optimization}, vol.~27, American Mathematical Soc., 2016.

\bibitem{van2018improvements}
{\sc J.~Van~Apeldoorn and A.~Gily{\'e}n}, {\em Improvements in quantum sdp-solving with applications}, arXiv preprint arXiv:1804.05058,  (2018).

\bibitem{van2017quantum}
{\sc J.~Van~Apeldoorn, A.~Gily{\'e}n, S.~Gribling, and R.~de~Wolf}, {\em Quantum sdp-solvers: Better upper and lower bounds}, in 2017 IEEE 58th Annual Symposium on Foundations of Computer Science (FOCS), IEEE, 2017, pp.~403--414.

\bibitem{vandenberghe1996semidefinite}
{\sc L.~Vandenberghe and S.~Boyd}, {\em Semidefinite programming}, SIAM Rev., 38 (1996), pp.~49--95.

\bibitem{vercellino2025hybridquantumclassicalbranchandpricemethod}
{\sc C.~Vercellino, M.~Y. Naghmouchi, W.~Coelho, G.~Vitali, A.~Scionti, P.~Viviani, O.~Terzo, and B.~Montrucchio}, {\em Hybrid quantum-classical branch-and-price method for the vertex coloring problem}, 2025, \url{https://arxiv.org/abs/2508.18887}.

\bibitem{wang2021polyhedral}
{\sc Y.~Wang, A.~Tanaka, and A.~Yoshise}, {\em Polyhedral approximations of the semidefinite cone and their application}, Comput. Optim. Appl., 78 (2021), pp.~893--913.

\bibitem{watts2023quantumsdp}
{\sc O.~Watts, Y.~Kikuchi, and L.~Coopmans}, {\em Quantum semidefinite programming with thermal pure quantum states}, 2023, \url{https://arxiv.org/abs/2310.07774}.

\bibitem{Westerheim2026DualVQE}
{\sc H.~Westerheim, J.~Chen, Z.~Holmes, I.~Luo, T.~Nuradha, D.~Patel, S.~Rethinasamy, K.~Wang, and M.~M. Wilde}, {\em Dual variational quantum eigensolver: A quantum algorithm to lower bound the ground-state energy}, Phys. Rev. A, 113 (2026), p.~032443, \url{https://doi.org/10.1103/twrt-y691}.

\bibitem{wolkowicz2012handbook}
{\sc H.~Wolkowicz, R.~Saigal, and L.~Vandenberghe}, {\em Handbook of semidefinite programming: theory, algorithms, and applications}, Springer Science \& Business Media, 2012.

\bibitem{wu2023inexact}
{\sc Z.~Wu, M.~Mohammadisiahroudi, B.~Augustino, X.~Yang, and T.~Terlaky}, {\em An inexact feasible quantum interior point method for linearly constrained quadratic optimization}, Entropy, 25 (2023), p.~330.

\bibitem{wu2025quantumIJOO}
{\sc Z.~Wu, P.~Sampourmahani, M.~Mohammadisiahroudi, and T.~Terlaky}, {\em A quantum dual logarithmic barrier method for linear optimization}, INFORMS J. Optim.,  (2025).

\end{thebibliography}

\end{document}